\documentclass{kapedbk} % Computer Modern font calls

\kluwerbib
\usepackage{epsf}
\def\gtrsim{\mathrel{\hbox{\rlap{\hbox{\lower4pt\hbox{$\sim$}}}\hbox{$>$}}}}
\let\ga=\gtrsim
\def\lesssim{\mathrel{\hbox{\rlap{\hbox{\lower4pt\hbox{$\sim$}}}\hbox{$<$}}}}
\let\la=\lesssim

\begin{document}
\articletitle{The Physics of Cluster Mergers}

\vskip-1.5truein
\noindent To appear in {\it Merging Processes in Clusters of Galaxies},
edited by L. Feretti, I. M. Gioia, and G. Giovannini
(Dordrecht: Kluwer), in press (2001)
\vskip0.85truein

% \articlesubtitle{This is an Article Subtitle}

\author{Craig L. Sarazin}
\affil{Department of Astronomy\\
University of Virginia}
\email{sarazin@virginia.edu}

% \and %% <<== will make `and' appear in the Table of Contents. Use
     %%      before the last author is listed.

%\author{Second Author}
%\affil{Author Affiliation\\
%Second Line of Affiliation}
%\email{secondauthor@anotheruniv.edu}

\begin{abstract}
Clusters of galaxies generally form by the gravitational merger of
smaller clusters and groups.
Major cluster mergers are the most energetic events in the Universe
since the Big Bang.
Some of the basic physical properties of mergers will be discussed,
with an emphasis on simple analytic arguments rather than numerical
simulations.
Semi-analytic estimates of merger rates are reviewed,
and a simple treatment of the kinematics of binary mergers is given.
Mergers drive shocks into the intracluster medium, and
these shocks heat the gas and should also accelerate
nonthermal relativistic particles.
X-ray observations of shocks can be used to determine the geometry and
kinematics of the merger.
Many clusters contain cooling flow cores;
the hydrodynamical interactions of these cores with the hotter, less
dense gas during mergers are discussed.
As a result of particle acceleration in shocks,
clusters of galaxies should contain very large populations of
relativistic electrons and ions.
Electrons with Lorentz factors $\gamma \sim 300$
(energies $E = \gamma m_e c^2 \sim 150$ MeV) are expected to
be particularly common.
Observations and models for the radio, extreme
ultraviolet, hard X-ray, and gamma-ray emission from nonthermal
particles accelerated in these mergers are described.
\end{abstract}

% ?? don't know what the set is
% \begin{keywords}
% \end{keywords}

\section{Introduction} \label{sec:intro}

Major cluster mergers are the most energetic events in the Universe
since the Big Bang.
Cluster mergers are the mechanism by which clusters are assembled.
In these mergers, the subclusters collide at velocities of
$\sim$2000 km/s,
releasing gravitational binding energies of as much as $\ga$$10^{64}$
ergs.
During mergers,
shocks are driven into the intracluster medium.
In major mergers, these hydrodynamical shocks dissipate energies of
$\sim 3 \times 10^{63}$ ergs;
such shocks are the major heating
source for the X-ray emitting intracluster medium.
The shock velocities in merger shocks are similar to those in
supernova remnants in our Galaxy, and we expect them to produce similar
effects.
Mergers shocks should heat and compress the X-ray emitting intracluster
gas, and increase its entropy.
We also expect that particle acceleration by these shocks will produce
relativistic electrons and ions, and these can produce synchrotron
radio, inverse Compton (IC) EUV and hard X-ray, and gamma-ray emission.

In this chapter, I will review some of the basic physics of cluster
mergers.
As later chapters discuss the optical, X-ray, and radio observations
of mergers, I will concentrate of theoretical issues.
Also, because later chapters discuss simulations of cluster mergers
and of large scale structure,
% (\nolinebreak\cite{Sch})
% and of large scale structure
% (\nolinebreak\cite{EG}),
I will mainly discuss analytical or semi-analytical aspects of
cluster mergers.
In \S~\ref{sec:basic_rates}, semi-analytic estimates of merger rates
based on Press-Schechter theory are reviewed.
Some simple estimates of the kinematics of binary cluster mergers
are given in
\S~\ref{sec:basic_kinematics}.
The thermal effects of merger shocks are discussed in
\S~\ref{sec:thermal},
with an emphasis on determining the physical conditions in mergers
from X-ray observations of temperatures and densities.
Many clusters and groups contain cooling flow cores.
During a merger, these cool cores will interact hydrodynamically with
the hotter, more diffuse intracluster gas
(\S~\ref{sec:cf}).
This can lead to the disruption of the cooling flow core,
as discussed in
\S~\ref{sec:cf_cflow}.
Recently, the Chandra X-ray Observatory has detected a number of
``cold fronts'' in merging clusters, which apparently are cool cores
moving through hot, shock heated, diffuse cluster gas
(\S~\ref{sec:cf_coldfront}).
Relativistic particles may be accelerated or reaccelerated in merger
shocks or turbulence generated by mergers.
The nonthermal effects of mergers are discussed in
\S~\ref{sec:nonthermal}.
The resulting radio, extreme ultraviolet, hard X-ray, and
gamma-ray emission is described.

\section{Basic Merger Rates and Kinematics}\label{sec:basic}

\subsection{Estimates of Merger Rates}\label{sec:basic_rates}

The rates of cluster mergers as a function of the cluster masses and
redshift can be estimated using a simple formalism originally proposed
by
Press \& Schechter (1974, hereafter PS),
and developed in more detail by
Bond et al.\ (1991)
and
Lacey \& Cole (1993),
among others.
Comparisons to observations of clusters and to numerical simulations
show that PS provides a good representation of the statistical
properties of clusters, if the PS parameters are carefully selected
(e.g., Lacey \& Cole 1993;
Bryan \& Norman 1998).
This formalism assumes that galaxies and clusters grow by the gravitational
instability of initially small amplitude gaussian density fluctuations
generated by some process in the early Universe.
The fluctuation spectrum is assumed to have larger amplitudes on smaller
scales.
Thus, galaxies and clusters form hierarchically, with lower mass objects
(galaxies and groups of galaxies) forming before larger clusters.
These smaller objects then merge to form clusters.

In the extended PS formalism, the density fluctuations in the Universe are
smoothed on a variety of mass scales.
Regions are assumed to collapse when their density exceeds a critical
value, which is usually taken to be the density for the collapse
for an isolated, spherical mass concentration of the same mass.
If one smooths the density fluctuations in some region on a variety of
mass scales, the average density may exceed the critical density for
collapse on a variety of different mass scales.
The assumption of the extended PS formalism is that material is
associated with the largest mass scale for which collapse has occurred,
and that smaller mass scales have merged into the larger object.
With these assumptions, the PS formalism allows one to estimate the
abundance of clusters as a function of their mass, and the rates
at which clusters merge.

Let $n(M,z) dM$ be the comoving number density of clusters with masses in
the range $M$ to $M + d M$ in the Universe at a redshift of $z$.
According to PS, the differential number density is given by
\begin{equation} \label{eq:PSdensity}
n(M,z) \, dM = \sqrt{ \frac{2}{\pi}} \, \frac{ \overline{\rho}}{M^{2}}
\, \frac{\delta_{c}(z)}{\sigma (M) } \,
\left| \frac{d \, \ln \, \sigma (M) }{d \, \ln \, M} \right|
\, \exp \left[- \frac{\delta_{c}^{2}(z)}{2\sigma^{2} (M) } \right]
\, dM
\, ,
\end{equation}
where $\overline{\rho}$ is the current mean density of the Universe,
$\sigma(M)$ is the current rms density fluctuation within a sphere of
mean mass $M$,
and $\delta_{c}(z)$ is the critical linear overdensity for a region to
collapse at a redshift $z$.

In Cold Dark Matter models, the initial spectrum of
fluctuations can be calculated for various cosmologies
(Bardeen et al.\ 1985).
Over the range of scales covered by clusters, it is generally sufficient
to consider a power-law spectrum of density perturbations,
which is consistent with these CDM models:
\begin{equation}  \label{eq:sigma}
\sigma (M) = \sigma_{8} \, \left( \frac{M}{M_{8}} \right) ^{-\alpha} \, ,
\end{equation}
where $\sigma_{8}$ is the present day rms density fluctuation on a scale of
8 $h^{-1}$ Mpc,
$M_8 = ( 4 \pi / 3 ) ( 8 \, h^{-1} \, {\rm Mpc} )^3 \bar{\rho}$
is the mass contained in a sphere of radius 8 $h^{-1}$ Mpc, 
and the Hubble constant is
$H_0 = 100 \, h$ km s$^{-1}$ Mpc$^{-1}$.
The exponent $\alpha$ is given by $\alpha = (n+3)/6$, where the power
spectrum of fluctuations varies with wavenumber $k$ as $k^n$.
The observations are generally reproduced with values of $-2 \la n \la -1$,
leading to $ 1/6 \la \alpha \la 1/3$.
The normalization of the power spectrum and overall present-day abundance of
clusters is set by $\sigma_8$.
The observed present-day abundance of clusters leads to
$ \sigma_8 \approx 0.6 \Omega_0^{-1/2}$,
where $\Omega_0 \equiv \bar{\rho} / {\rho_c} $ is the ratio of the current
mass density to the critical mass density, $\rho_c = 3 H_0^2 / ( 8 \pi G )$
(e.g., Bahcall \& Fan 1998).

The evolution of the density of clusters is
encapsulated in the critical over-density $\delta_{c}(z)$ in
equation~(\ref{eq:PSdensity}).
In general, $\delta_c (z) \propto 1/D(t)$, where
$D(t)$ is the growth factor of linear perturbations as a function of
cosmic time  $t$
(see Peebles [1980], \S~11 for details).
Expressions for the $\delta_c (z)$ in different cosmological models
are:
\begin{equation}   \label{eq:delta}
\delta_{c}(z) = \left\{
\begin{array}{ll} 
\frac{3}{2} D( t_0 ) \left[ 1 + 
\left( \frac{t_{\Omega}}{t}
\right)^{\frac{2}{3}}
\right]   &
(\Omega_0 < 1 \, , \, \Omega_\Lambda = 0) \\
\frac{3(12\pi)^\frac{2}{3}}{20} \left(
\frac{t_0 }{t}
\right)^{\frac{2}{3}} &
(\Omega_0  = 1 \, , \, \Omega_\Lambda = 0) \\
\frac{D(t_0 )}{D(t)} 
\left(\frac{3(12\pi)^\frac{2}{3}}{20}\right) 
\left(1 + 0.0123 \ \log \Omega_{z}
\right) &
(\Omega_0 + \Omega_{\Lambda} = 1)
\end{array}
\right. 
\end{equation}
Here, $\Omega_{\Lambda}$ gives the contribution due to a cosmological
constant $\Lambda$, where
$\Omega_{\Lambda} \equiv \Lambda / ( 3 H_0^2 )$.
For the open model ($\Omega_0 < 1$, $\Omega_\Lambda = 0$),
$t_{\Omega} \equiv \pi H_0^{-1} \Omega_0$
$\left( 1 - \Omega_0 \right)^{-\frac{3}{2}}$
represents the epoch at which a nearly constant expansion takes over
and no new clustering can occur.
The growth factor can be expressed as
\begin{equation}  \label{eq:growth_factor}
D(t)=\frac{3\ \sinh \eta \left( \sinh \eta - \eta \right)}
	  {\left( \cosh \eta - 1 \right)^{2}} - 2
\end{equation} 
where $\eta$ is the standard parameter in the cosmic expansion equations
(Peebles 1980, eqn.~13.10)
\begin{equation} \label{eq:eta}
\begin{array}{ll}
\frac{1}{1+z} = \frac{\Omega_0}
{2 \left( 1 - \Omega_0 \right)}
\left( \cosh \eta - 1 \right) \, , \\
H_0 t = \frac{\Omega_0 }
{2 \left( 1 - \Omega_0 \right)^{\frac{3}{2}}}
\left( \sinh \eta - \eta \right) \, .
\end{array}
\end{equation}
The solution for $\delta_{c}$ in the Einstein-deSitter model
($\Omega_0  = 1$, $\Omega_\Lambda = 0$)
can be obtained from the open model solution by the limit
$t_{\Omega}/t \rightarrow \infty$.
The expression for $\delta_{c}$ in the flat model
$(\Omega_0 + \Omega_{\Lambda} = 1$)
is an approximation given by
Kitayama \& Suto (1996).
Here $\Omega_{z}$ is the value of the mass density ratio $\Omega$ at
the redshift $z$,
\begin{equation} \label{eq:omega_f}
\Omega_{z} = \frac{\Omega_0
	\left( 1+z \right)^{3}}
	{\Omega_0 \left( 1+z \right)^{3} + \Omega_{\Lambda}} \, .
\end{equation}
In this model the growth factor can be written as
\begin{equation} \label{eq:growth_flat}
D(x)= \frac{(x^{3} + 2)^{1/2}}{x^{3/2}} \,
\int_{0}^{x} x^{3/2} \, (x^{3}+2)^{-3/2}dx
\end{equation}
(Peebles 1980, eqn.~13.6) where
$x_0 \equiv ( 2 \Omega_{\Lambda} / \Omega_0 )^{1/3}$ and $x = x_0/(1+z)$.

The PS formalism also provides estimates of the merger history,
rates, and probabilities for clusters.
For example, the probability that a cluster with a mass
$M_0$ at the present time $t_0$ had a progenitor with a mass of $M$ at
an earlier time $t < t_0$ is given by
\begin{eqnarray}
\frac{d p}{d M} ( M , t | M_0, t_0 ) & = &
\frac{ \delta_c ( t ) - \delta_c ( t_0 ) }{ \sqrt{2 \pi}
\left[ \sigma^2 ( M ) - \sigma^2 ( M_0 ) \right]^{3/2} } \,
\left( \frac{ M_0 }{ M } \right) \,
\left| \frac{ d \, \sigma^2 ( M ) }{ d \, M } \right| \nonumber \\
& & \qquad \exp \left\{ - \frac{
\left[ \delta_c ( t ) - \delta_c ( t_0 ) \right]^2 }{ 2
\left[ \sigma^2 ( M ) - \sigma^2 ( M_0 ) \right] } \right\} \, .
\label{eq:merger_progentor}
\end{eqnarray}
Similarly, the probability that a cluster of mass $M$
undergoes a merger with cluster of mass $\Delta M$ per unit time
is given by
\begin{eqnarray}
\frac{d^2 p}{d \Delta M \, dt} & = &
\sqrt{ \frac{2}{\pi}} \,
\frac{\delta_{c}(z)}{\sigma (M' ) } \,
\left| \frac{d \, \ln \delta_{c}(z)}{d t} \right| \,
\left| \frac{d \, \ln \, \sigma (M' ) }{d \, M'} \right| \,
\left[ 1 - \frac{\sigma^2 ( M' )}{\sigma^2 ( M )} \right]^{-3/2} \nonumber \\
& & \qquad \times \exp \left\{ - \frac{\delta_{c}^{2}(z)}{2} \,
\left[ \frac{1}{\sigma^{2} ( M' )} - \frac{1}{\sigma^{2} ( M )} \right]
\right\} \, ,
\label{eq:PSmerger}
\end{eqnarray}
where $M' = M + \Delta M$.

\begin{figure}[t]
\vskip3.7truein
% \special{psfile=merger_tree.ps hscale=50 vscale=50 hoffset=24
\includegraphics{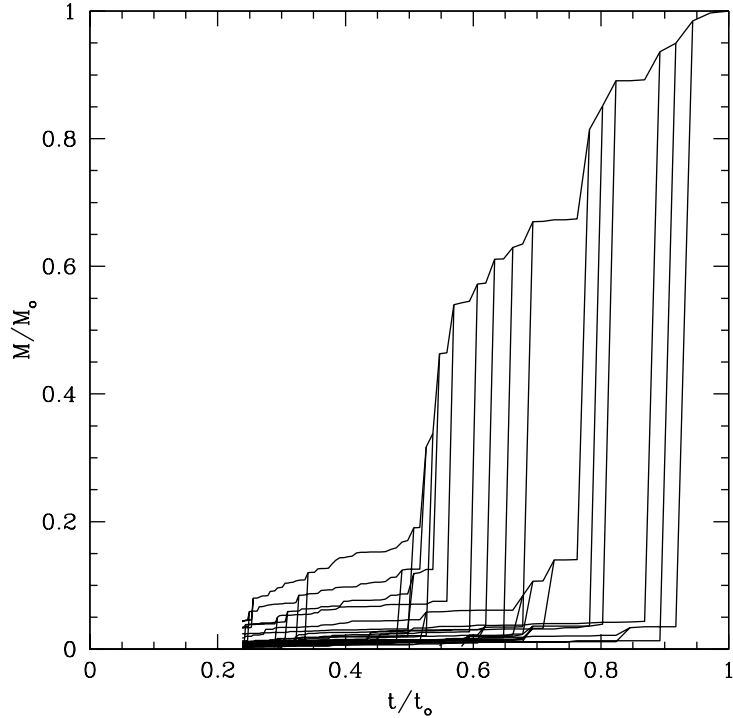}
\caption[]{An example of a PS merger tree for a cluster of galaxies
with a final mass of $M_0 = 10^{15} h^{-1} \, M_\odot$
(Randall \& Sarazin 2001).
The mass is shown as a function of the age of the Universe $t$;
the present age is $t_0$.
This model was for an open Universe with $\Omega_0 = 0.3$ and
$\Omega_\Lambda = 0$.
\label{fig:merger_tree}}
\end{figure}

These probability distributions can be used to make Monte Carlo simulations
of the merger histories which produced clusters of a various masses
at present.
Figure~\ref{fig:merger_tree} shows one such ``merger tree'' for a
cluster with a mass of $10^{15}$ $M_\odot$ at the present time
(Randall \& Sarazin 2001).
At least to the extent that the development of a cluster can be
treated as a series of separate, discrete merger events separated by
periods of approximate equilibrium
(the ``punctuated equilibrium'' model;
Cavaliere, Menci, \& Tozzi 1999),
\nocite{CMT}
these merger histories can be used to determine the effects of mergers
on clusters.

\subsection{Estimates of Merger Kinematics}\label{sec:basic_kinematics}

\begin{figure}[t]
\vskip2.6truein
% \vskip3.8truein
% \epsfbox{merger_kinematics.ps}
% \special{psfile=merger_kinematics.ps hscale=100 vscale=100
\includegraphics{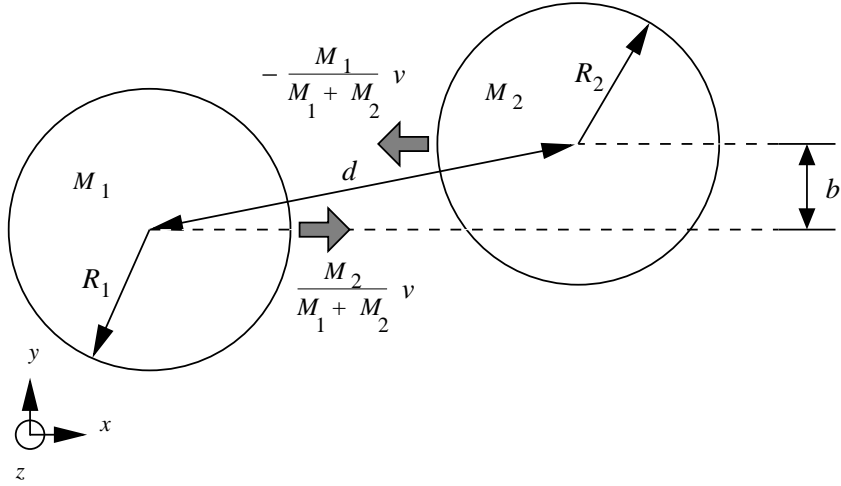}
% \special{psfile=merger_kinematics.ps hscale=100 vscale=100 hoffset=-130
% voffset=-80}
\caption[]{A schematic diagram of the kinematics for a merger between
two subclusters of masses $M_1$ and $M_2$ and radii $R_1$ and $R_2$.
The separation of the cluster centers is $d$, and the impact parameter is
$b$, and the initial relative velocity is $v$.
\label{fig:merger_kinematics}}
\end{figure}

I now give some simple analytic argments to estimate the kinematics
of an individual binary merger collision.
The kinematic quantities describing the merger are defined in
Figure~\ref{fig:merger_kinematics},
which is taken from Ricker \& Sarazin (2001).
The two subclusters have masses $M_1$ and $M_2$.
Let $d$ be the separation of the centers of the two subclusters,
let $v$ be the relative velocity of the centers, and let
$b$ be the impact parameter of the collision.

\subsubsection{Turn-Around Distances}\label{sec:basic_kinematics_d0}

Assume that the two subclusters of mass $M_1$ and $M_2$ merge 
at some time $t_{\rm merge}$ (the age of the Universe at the time of
the merger).
It is assumed that the two subclusters have fallen together
from a large distance $d_0$ with (possibly) nonzero angular momentum.
(The exact value of $d_0$ does not affect the collision velocity very strongly
as long as it is large and the infall velocity approaches free-fall
from infinity.)
For the purpose of computing the initial relative velocity,
we approximate the two clusters as point masses.
We assume that the two subclusters were initially expanding away from
one another in the Hubble flow, and that their radial velocity was zero
at their greatest separation $d_0$.
If we assume that the two subclusters dominate the mass in the region
of the Universe they occupy, we can treat their initial expansion
and recollapse as the orbit of two point masses, and Kepler's
Third Law gives the greatest separation as
\begin{eqnarray}
d_0 & \approx & \left[ 2 G \left( M_1 + M_2 \right) \right]^{1/3}
\left( \frac{t_{\rm merge}}{\pi} \right)^{2/3} \nonumber \\
& \approx & 4.5
\left( \frac{M_1 + M_2}{10^{15} \, M_\odot} \right)^{1/3}
\left( \frac{t_{\rm merge}}{10^{10} \, {\rm yr}} \right)^{2/3} 
\, {\rm Mpc} \, .
\label{eqn:init sep}
\end{eqnarray}

\subsubsection{Merger Velocities}\label{sec:basic_kinematics_v}

At the separation $d_0$, the clusters are assumed to have zero relative
radial velocity; hence their orbital angular momentum and energy are
\begin{eqnarray}
J_{\rm orb} &\approx& m v_0 d_0 \nonumber \\
E_{\rm orb} &\approx& \frac{1}{2} m v_0^2 - \frac{G M_1 M_2}{d_0} \ ,
\label{eqn:init1}
\end{eqnarray}
where their reduced mass is
\begin{equation} \label{eqn:reduced mass}
m \equiv {M_1M_2\over M_1+M_2}\ ,
\end{equation}
and $v_0$ is their initial relative transverse velocity.
At the separation $d$, the relative velocity $v$ is perpendicular
to the direction of $b$, so we can write
\begin{eqnarray}
J_{\rm orb} &\approx& m v b \nonumber \\
E_{\rm orb} &\approx& \frac{1}{2} m v^2 - \frac{G M_1 M_2}{d} \ .
\label{Eqn:ang mom and energy}
\end{eqnarray}
Conserving angular momentum and energy, we eliminate $v_0$ and find
\begin{equation} \label{eqn:initial velocity2}
v^2 \approx 2G\left(M_1+M_2\right)\left({1\over d}-{1\over d_0}\right)
      \left[1-\left(b\over d_0\right)^2\right]^{-1}\ ,
\end{equation}
or
\begin{equation} \label{eqn:initial velocity}
v \approx 2930
\left( \frac{M_1 + M_2}{10^{15} \, M_\odot} \right)^{1/2}
\left( \frac{d}{1 \, {\rm Mpc}} \right)^{-1/2}
\left[ \frac{1 - \frac{d}{d_0}}{1-\left( \frac{b}{d_0} \right)^2}
\right]^{1/2} \, {\rm km \, s}^{-1} \, .
\end{equation}

\subsubsection{Angular Momenta, Impact Parameters, and Transverse Velocities}
\label{sec:basic_kinematics_b}

The remaining kinematic parameter for the merger is the impact
parameter $b$, or equivalently the orbital angular momentum $J_{\rm orb}$ or
the initial tangential velocity $v_0$.
In principal, a range of values are possible for mergers of subclusters
with similar masses and similar merger epochs $t_{\rm merge}$.
The angular momentum will be determined by tidal torques from surrounding
material.
Thus, I give a estimate of the range of possible values based on
the linear-theory result for the dimensionless spin of dark-matter
halos;
this argument is given in Ricker \& Sarazin (2001).
The spin parameter $\lambda$ is defined as
(Peebles 1969)
\begin{equation}
\label{eq:dimensionless spin}
\lambda \equiv {J|E|^{1/2}\over GM^{5/2}}\ .
\end{equation}
Here $J$ is the total angular momentum of the halo, 
$E$ is its total energy,
and $M$ is its mass.
In linear theory, the average value of $\lambda$ is expected to be
approximately constant, independent of the mass of the halo.
Recently, Sugerman, Summers, and Kamionkowski (2000) have performed a
\nocite{SSK}
detailed
comparison of linear-theory predictions to actual angular momenta of
galaxies formed in cosmological $N$-body/hydro calculations.
These simulations did not include cooling or star formation, so at the
upper end of the mass range they studied their results should carry over
to clusters.
They find, in agreement with
White (1984),
that linear theory overpredicts
the final angular momentum of galaxies by roughly a factor of three, with
a large ($\sim 50\%$) dispersion in the ratio of the linear-theory
prediction to the actual value.
However, given the uncertainties, the angular momenta agree with the
results in equation~(\ref{eq:dimensionless spin}) for a value of
$\lambda \approx 0.05$.
Thus, we will assume that the average total angular momenta of clusters of
galaxies are given by
\begin{equation}
\label{eq:angular momentum}
J \approx \frac{\lambda G M^{5/2}}{|E|^{1/2}}\ ,
\end{equation}
with $\lambda \approx 0.05$.
The normal virial relations for clusters imply that the energies
of clusters scale with their mass as $| E | \propto M^{5/3}$,
which implies that the angular momenta scale as
$ J \propto M^{5/3}$ as well.

Let us take the halo to be the final merged cluster.
Its final total angular momentum is the sum of the angular momenta of
the two subclusters plus the orbital angular momentum $J_{\rm orb}$.
Applying equation~(\ref{eq:angular momentum}) to the initial masses
$M_1$ and $M_2$ and the final mass $M_1 + M_2$ and taking the difference
gives the orbital angular momentum $J_{\rm orb}$.
We assume that the angular momenta are correlated (i.e., that they lie
along the same direction), since they are all produced by approximately
the same local tidal field.
The final energy of the merged cluster is the sum of the energies
of the initial subclusters plus the orbital energy $E_{\rm orb}$.
The rotational kinetic energies can be ignored as they are only
a fraction $\sim 2 \lambda^2 \la 1$\% of the total energies.

Using these relations, the average orbital angular momentum of the merger is
found to be
\begin{equation}
\label{eq:Jorb}
J_{\rm orb} \approx \frac{\lambda G M_1 M_2}{
\left[ \frac{G ( M_1 + M_2 )}{d_0} - \frac{1}{2} v_0^2 \right]^{1/2}}
\,
f(M_1,M_2) \, .
\end{equation}
Here, the function $f(M_1, M_2 )$ corrects for the internal angular momenta
and energy of the subclusters.
This correction can be written as
\begin{equation}
\label{eq:fcorr}
f( M_1 , M_2 ) \equiv \frac{ ( M_1 + M_2 )^3}{ M_1^{3/2} M_2^{3/2}} \,
\left[ 1 -
\frac{ \left( M_1^{5/3} + M_2^{5/3} \right)}{\left( M_1 + M_2
\right)^{5/3}}
\right]^{3/2} \, ,
\end{equation}
but it only depends on the ratio $( M_< / M_> )$ of the smaller to
larger mass of the two subclusters.
It varies between
$4( 2^{2/3} - 1 )^{3/2} \approx 1.80 \le f(M_1, M_2) \le (5/3)^{3/2} \approx
2.15$, so that $f( M_1, M_2 ) \approx 2$.
The kinetic energy term $v_0^2 / 2$ in the denominator of
equation~(\ref{eq:Jorb}) can be shown to be approximately
$2 \lambda^2 \approx 1$\% of the potential energy term.
Thus, this term can be dropped to yield
\begin{equation}
\label{eq:Jorb2}
J_{\rm orb} \approx \lambda M_1 M_2 \,
\sqrt{ \frac{G d_0}{M_1 + M_2} } \,
f(M_1,M_2) \, .
\end{equation}

The corresponding initial transverse velocity is
\begin{eqnarray}
v_0 & \approx & \lambda \,
\sqrt{ \frac{G ( M_1 + M_2) }{d_0}} \,
f(M_1,M_2) \nonumber \\
& \approx & 93
\left( \frac{\lambda }{0.05} \right) \,
\left( \frac{M_1 + M_2}{10^{15} \, M_\odot} \right)^{1/2}
\left( \frac{d_0}{5 \, {\rm Mpc}} \right)^{-1/2} \,
\left( \frac{f}{2} \right) \,
{\rm km \, s}^{-1} \, .
\label{eq:v0}
\end{eqnarray}

After the clusters have fallen towards one another to a separation $d$,
the impact parameter for the collision is
(Figure~\ref{fig:merger_kinematics})
\begin{equation}
b \approx \left( \frac{v_0}{v} \right) \, d_0 \, ,
\label{eq:b}
\end{equation}
where the infall velocity is given by 
equation~(\ref{eqn:initial velocity}).
Note that equation~(\ref{eq:b}) implies that $b \ll d_o$, so that one
can drop the $( b / d_0 )$ term in equation~(\ref{eqn:initial velocity}).
Substituting equations~(\ref{eqn:initial velocity}) \& (\ref{eq:v0})
into equation~(\ref{eq:b}) gives
\begin{eqnarray}
b ~~ \approx & \lambda \, \sqrt{ \frac{d_0 d}{2}} \,
\left( 1 - \frac{d}{d_0} \right)^{-1/2} \,
f(M_1,M_2) ~~~~~~~~~~~~~~~~~~~~~~~~~~~~~ \nonumber \\
 ~~ \approx & 160
\left( \frac{\lambda}{0.05} \right)
\left( \frac{d}{1 \, {\rm Mpc}} \right)^{1/2}
\left( \frac{d_0}{5 \, {\rm Mpc}} \right)^{1/2}
\left( 1 - \frac{d}{d_0} \right)^{-1/2}
% \left( \frac{d_0}{5 \, {\rm Mpc}} \right)^{1/2} \nonumber \\
% & & \qquad \times \left( 1 - \frac{d}{d_0} \right)^{-1/2}
\left( \frac{f}{2} \right)
{\rm kpc} \, .
\label{eq:b2}
\end{eqnarray}

Thus, most mergers are expected to involve fairly small impact parameters,
comparable to the sizes of the gas cores in clusters.
Many examples are known of mergers where the X-ray morphology
suggests a small offset;
an example is the merger in the cluster surrounding Cygnus-A
(Markevitch, Sarazin, \& Vikhlinin 1999).
\nocite{MSV}
However, the preceding arguments are approximate and statistical,
and mergers with larger impact parameters are also expected to occur;
based on the X-ray image and temperature map, it is likely that Abell~3395
is an example of such a merger
(\nolinebreak\cite{MFSV}).
Larger impact parameters may occur in mergers involving more than two
subclusters.
On the other hand, the distribution of impact parameters may be biased to
lower values if most mergers occur along large scale structure filaments
(e.g., \cite{EG}).

\section{Thermal Physics of Merger Shocks}\label{sec:thermal}

The intracluster medium (ICM) is generally close to hydrostatic equilibrium
in clusters which are not undergoing strong mergers.
The virial theorem then implies that the square of the thermal velocity
(sound speed) of the ICM is comparable to the gravitational potential.
During a merger, the infall velocities of the subclusters
(equation~\ref{eqn:initial velocity}) are comparable to the escape
velocity, which implies that the square of the infall velocity is larger
(by roughly a factor of two) than the gravitational potential.
Thus, the motions in cluster mergers are expected to be supersonic,
but only moderately so.
As a result, one expects that cluster mergers will drive shock waves into
the intracluster gas of the two subclusters.
Let $v_s$ be the velocity of such a shock wave relative to the preshock
intracluster gas.
The sound speed in the preshock gas is
$c_s = \sqrt{ (5/3) P / \rho}$, where $P$ is the gas pressure and $\rho$ is
the density.
Then, the Mach number of the shock is
${\cal M} \equiv v_s / c_s $.
Based on the simple argument given above and confirmed by merger simulations
(\nolinebreak\cite{SM};
\cite{RSB};
\cite{RiS};
\cite{Sch}),
one expects shocks with Mach numbers of ${\cal M} \la 3$.
Stronger shocks may occur under some circumstances, such as in the outer
parts of clusters, or in low mass subclusters merging with more massive
clusters.
However, in the latter case, the shocks in the less massive subcluster
may also be weak if the intergalactic gas in the smaller subcluster is
denser than that in the more massive subclusters
(\S~\ref{sec:cf}).

% Merger shocks heat and compress the intracluster gas, and these effects
% can be used to determine the geometry and kinematics of the merger.

Shocks are irreversible changes to the gas in clusters, and thus increase
the entropy $S$ in the gas.
A useful quantity to consider is the specific entropy per particle in the
gas, $s \equiv S / N$, where $N$ is the total number of particles.
To within additive constants, the specific entropy of an ideal gas is
\begin{eqnarray}
s & = & \frac{3}{2} \, k \,
\ln \left( \frac{P}{\rho^{5/3}} \right)
\, ,\nonumber \\
& = & \frac{3}{2} \, k \,
\ln \left( \frac{T}{\rho^{2/3}} \right) \, ,
\label{eq:entropy}
\end{eqnarray}
where $T$ is the gas temperature.
Observations of X-ray spectra can be used to determine $T$, while the X-ray
surface brightness depends on $\rho^2$.
Thus, one can use X-ray observations to determine the specific entropy in
the gas just before and just after apparent merger shocks seen in the
X-ray images.
Since merger shocks should produce compression, heating, pressure
increases, and entropy increases, the corresponding increase in all
of these quantities (particularly the entropy) can be used to check
that discontinuities are really shocks
(e.g., not ``cold fronts'' or other contact discontinuities,
\S~\ref{sec:cf_coldfront}).

Markevitch et al.\ (1999) applied this test to ASCA
temperature maps and ROSAT images of Cygnus-A and Abell~3667,
two clusters which appeared to show strong merger shocks.
\nocite{MSV}
(Recent Chandra images have cast doubt on the interpretation of
Abell~3667
[Vikhlinin, Markevitch, \& Murray 2001b].)
\nocite{VMMb}
In Cygnus-A, the increase in specific entropy in the shocked regions is
roughly $ \Delta s \approx (3/2) k$.
The specific heat per particle $q$ which must be dissipated to produce
this change in entropy is $q \approx T \Delta s \approx (3/2) k T$, or
about the present specific heat content in the shocked gas.
Thus, these observations provide a direct confirmation that merger shocks
contribute significantly to the heating of the intracluster gas.

% ?? more here

% ?? more on numerical simualtions

\subsection{Shock Kinematics}\label{sec:thermal_kinematics}

The variation in the hydrodynamical variables in the intracluster medium
across a merger shock are determined by the standard
Rankine--Hugoniot jump conditions
(e.g., \cite{LanLif}, \S~85),
if one assumes that all of the dissipated shock energy is thermalized.
Consider a small element of the surface of a shock (much smaller than
the radius of curvature of the shock, for example).
The tangential component of the velocity is continuous at the shock,
so it is useful to go to a frame which is moving with that element of
the shock surface, and which has a tangential velocity which is equal
to that of the gas on either side of the shock.
In this frame, the element of the shock surface is stationary, and the
gas has no tangential motion.
Let the subscripts 1 and 2 denote the preshock and postshock gas;
thus, $v_1 = v_s$ is the longitudinal velocity of material into the shock
(or alternative, the speed with which the shock is advancing into the
preshock gas).
Conservation of mass, momentum, and energy then implies the following
jump conditions
\begin{eqnarray}
\rho_1 v_1 & = & \rho_2 v_2 \, , \nonumber \\
P_1 + \rho_1 v_1^2 & = & P_2 + \rho_2 v_2^2 \, ,\nonumber \\
w_1 + \frac{1}{2} v_1^2 & = & w_2 + \frac{1}{2} v_2^2 \, .
\label{eq:jump}
\end{eqnarray}
Here, $w = P/\rho + \epsilon$ is the enthalpy per unit mass in the gas,
and $\epsilon$ is the internal energy per unit mass.
If the gas behaves as a perfect fluid on each side of the shock, the
internal energy per unit mass is given by
\begin{equation}
\epsilon = \frac{1}{\gamma_{\rm ad} - 1} \, \frac{P}{\rho} \, ,
\label{eq:int_energy}
\end{equation}
where $\gamma_{\rm ad}$ is the ratio of specific heats (the adiabatic index)
and is $\gamma_{\rm ad} = 5/3$ for fully ionized plasma.
The jump conditions can be rewritten as:
\begin{eqnarray}
\frac{P_2}{P_1} & = &
\frac{ 2 \gamma_{\rm ad}}{\gamma_{\rm ad} + 1} {\cal M}^2 -
\frac{\gamma_{\rm ad} - 1}{\gamma_{\rm ad} + 1} \, \nonumber \\
\frac{v_2}{v_1} =
\frac{\rho_1}{\rho_2} \equiv \frac{1}{C} & = &
\frac{ 2 }{\gamma_{\rm ad} + 1} \frac{1}{{\cal M}^2} +
\frac{\gamma_{\rm ad} - 1}{\gamma_{\rm ad} + 1} \, ,
\label{eq:jumpM}
\end{eqnarray}
where $C \equiv \rho_2 / \rho_1$ is the shock compression.

If one knew the velocity structure of the gas in a merging cluster, one
could use these jump condition to derive the temperature, pressure, and
density jumps in the gas.
At present, the best X-ray spectra for extended regions in clusters of
galaxies have come from CCD detectors on ASCA, Chandra, and XMM/Newton.
CCDs have a spectral resolution of $>$100 eV at the Fe K line at
7 keV, which translates into a velocity resolution of $>$4000 km/s.
Thus, this resolution is (at best) marginally insufficient to measure
merger gas velocities in clusters.
In a few cases with very bright regions and simple geometries, the
grating spectrometers on Chandra and especially XMM/Newton may be useful.
However, it is likely that the direct determinations of gas velocities in
most clusters will wait for the launch of higher spectral resolution
nondispersive spectrometers on Astro-E2 and Constellation-X.

At present, X-ray observations can be used to directly measure the
temperature and density jumps in merger shocks.
Thus, one needs to invert the jump relations to give the merger shock
velocities for a given shock temperature, pressure, and/or density increase.
If the temperatures on either side of the merger shock can be measured
from X-ray spectra, the shock velocity can be inferred from
(\nolinebreak\cite{MSV})
\begin{equation}
\Delta v_{s} = \left[\frac{k T_1}{\mu m_p}\left(C-1\right)\left(
\frac{T_2}{T_1}-\frac{1}{C}\right)
\right]^{1/2} \, ,
\label{eq:jumpv}
\end{equation}
where $ \Delta v_{s} = v_1 - v_2 = [(C-1)/C] v_s$ is the velocity
change across the shock, and
$\mu$ is the mean mass per particle in units of the proton mass $m_p$.
The shock compression $C$ can be derived from the temperatures as
\begin{equation}
\frac{1}{C} =
\left[\frac{1}{4}\left(\frac{\gamma_{\rm ad}+1}{\gamma_{\rm ad}-1}\right)^2
\left(\frac{T_2}{T_1}-1\right)^2 +\frac{T_2}{T_1}\right]^{1/2}
-\frac{1}{2}\frac{\gamma_{\rm ad}+1}{\gamma_{\rm ad}-1}\left(\frac{T_2}{T_1}-1\right) \, .
\label{eq:jumpC}
\end{equation}
Alternatively, the shock compression can be measured directly from
the X-ray image.
However, it is difficult to use measurements of the shock compression alone
to determine the shock velocity, for two reasons.
First, a temperature is needed to set the overall scale of the
velocities;
as is obvious from equation~(\ref{eq:jumpM}), the shock compression allows
one to determine the Mach number $\cal M$ but not the shock velocity.
The second problem is that temperature or pressure information is needed
to know that a discontinuity in the gas density is a shock, and not
a contact interface
(e.g., the ``cold fronts'' discussed in \S~\ref{sec:cf_coldfront} below).

X-ray temperature maps of clusters have been used to derive the merger
velocities using these relations.
Markevitch et al.\ (1999) used ASCA observations to determine the
kinematics of mergers in three clusters
(Cygnus-A, Abell~2065, and Abell~3667).
Because of the poor angular resolution of ASCA, these analyses were
quite uncertain.
More recently, possible shocks have been detected in Chandra images of a number
of merging clusters
(e.g., Abell~85,
Kempner, Sarazin, \& Ricker 2001;
\nocite{KSR}
Abell~665, \cite{MVMV};
Abell~3667, \cite{VMMb}),
and the shock jump conditions have been applied to determine the kinematics
in these clusters.

The simplest case is a head-on symmetric merger
($b = 0$ and $M_1 = M_2$) at an early stage when the shocked region lies
between the two cluster centers.
Markevitch et al.\ (1999) suggest that the Cygnus-A cluster is an example.
If the gas within the shocked region is nearly stationary, then
the merger velocity of the two subclusters is just
$v = 2 \Delta v_s$.
Applying these techniques to the ASCA temperature map for the Cygnus-A
cluster,
Markevitch et al.\ found a merger velocity of
$v \approx 2200$ km/s.
This simple argument is in reasonable agreement with the results of
numerical simulations of this merger
(\nolinebreak\cite{RiS}).
The radial velocity distribution of the galaxies in this cluster
is bimodal
(\nolinebreak\cite{OLMH}),
and consistent with a merger velocity of $\sim$2400 km/s.

One can compare the merger velocities derived from the temperature jumps
in the merger shocks with the values predicted by free-fall from the
turn-around radius
(equation~\ref{eqn:initial velocity}).
In the case of Cygnus-A,
Markevitch et al.\ (1999) found good agreement with the
the free-fall velocity of $\sim$2200 km/s.
This consistency suggests that the shock energy is effectively thermalized,
and that a major fraction does not go into turbulence, magnetic fields,
or cosmic rays. 
Thus, the temperature jumps in merger shocks can provide an important
test of the relative roles of thermal and nonthermal processes in
clusters of galaxies.
Further tests should be possible by comparing shock heating with velocities
determined from optical redshifts, from direct velocity measurements in the gas
with Astro-E2 and Constellation-X, and from infall arguments.

\subsection{Nonequilibrium Effects} \label{sec:thermal_nonequil}

Cluster mergers are expected to produce collisionless shocks, as occurs
in supernova remnants.
As such, nonequilibrium effects are expected, including nonequipartition
of electrons and ions and nonequilibrium ionization
(Markevitch et al.\ 1999;
\cite{Tak99},2000).
\nocite{Tak00}
Collisionless shocks are generally not as effective in heating electrons
as ions.
Assuming that the postshock electrons are somewhat cooler than the ions,
the time scale for electron and protons to approach equipartition as
a result of Coulomb collisions in a hot ionized gas is
(\nolinebreak\cite{spitzer})
\begin{eqnarray}
t_{\rm eq} & = &
\frac{3 m_p m_e}{8 \sqrt{2 \pi} n_e e^4 \ln \Lambda} \,
\left( \frac{ k T_e }{ m_e } \right)^{3/2} \nonumber \\
& \approx & 2.1 \times 10^8 \,
\left( \frac{ T_e }{ 10^8 \, {\rm K} } \right)^{3/2} \,
\left( \frac{ n_e }{ 0.001 \, {\rm cm}^{-3} } \right)^{-1} \,
{\rm yr} \, ,
\label{eq:equipartition}
\end{eqnarray}
where $n_e$ and $T_e$ are the electron number density and temperature,
respectively, and
$\Lambda$ is the Coulomb factor.
The relative velocity between the postshock gas and the shock front is
(1/4)$v_s$;
thus, one would expect the electron temperature to reach equipartition a
distance of
\begin{equation} \label{eq:deq}
d_{\rm eq} \approx 160
\left( \frac{ v_s }{ 3000 \, {\rm km/s} } \right) \,
\left( \frac{ T_e }{ 10^8 \, {\rm K} } \right)^{3/2} \,
\left( \frac{ n_e }{ 0.001 \, {\rm cm}^{-3} } \right)^{-1} \,
{\rm kpc}
\end{equation}
behind the shock front.
Of course, it is the electron temperature (rather than the ion or average
temperature) which determines the shape of the X-ray spectrum.
This distance is large enough to insure that the lag could be spatially
resolved in X-ray observations of low redshift clusters.
Similar effects might be expected through non-equilibrium ionization.

On the other hand, it is likely that the nonequilibrium effects in cluster
merger shocks are much smaller than those in supernova blast wave
shocks because of the low Mach numbers of merger shocks.
That is, the preshock gas is already quite hot
(both electrons and ions) and highly ionized.
Moreover, a significant part of the heating in low Mach number shocks is
due to adiabatic compression, and this would still act on the electrons
in the postshock gas in merger shocks, even if there were no collisionless
heating of electrons.
For example, in a ${\cal M} = 2$, $\gamma_{\rm ad} = 5/3$ shock, the
total shock increase in temperature is a factor of 2.08
(eq.~\ref{eq:jumpM}).
The shock compression is $C = 2.29$, so adiabatic compression increases
the electron temperature by a factor of $C^{2/3} = 1.74$,
which is about 83\% of the shock heating.

\section{Mergers and Cool Cluster Cores}\label{sec:cf}

\subsection{Cooling Flows vs.\ Mergers}\label{sec:cf_cflow}

The centers of a significant fraction of clusters of galaxies
have luminous cusps in their X-ray surface brightness
known as ``cooling flows''
(see \cite{F94} for an extensive review).
In every case, there is a bright (cD) galaxy at the center of the
cooling flow region.
% Although there is a great deal of interesting physics associated with
% these cooling flows, I concentrate here on the basic observational aspects
% related to mergers.
The intracluster gas densities in these regions are much higher than the
average values in the outer portions of clusters.
X-ray spectra indicate that there are large amounts of gas at
low temperatures (down to $\sim$$10^7$ K), which are much cooler than those
in the outer parts of clusters.
The high densities imply rather short cooling times $t_{\rm cool}$
(the time scale for the gas to cool to low temperature due to its own
radiation).
The hypothesis is that the gas in these regions is cooling from higher
intracluster temperature ($\sim$$10^8$ K) down to these lower
temperatures as a result of the energy loss due to the X-ray emission
we observe.
Typical cooling rates are $\sim$100 $M_\odot$ yr$^{-1}$.
The cooling times, although much shorter than the Hubble time, are
generally much longer than the dynamical (i.e., sound crossing time) of the
gas in these regions.
As a result, the gas is believed to remain nearly in hydrostatic equilibrium.
Thus, the gas must compress as it cools to maintain a pressure which can
support the weight of the overlying intracluster medium.

The primary observational characteristics of cooling flows are
very bright X-ray surface brightnesses which increase rapidly toward the
center of the cluster.
The high surface brightnesses imply high gas densities which also
increase rapidly towards the cluster center.
These regions contain cooler cluster gas.

Empirically, there is significant indirect evidence that mergers
disrupt cooling flows.
There is a strong statistical anticorrelation between cooling flows
and/or cooling rates, 
and irregular structures in clusters as derived by statistical analysis
of their X-ray images
(\nolinebreak\cite{BT}).
The irregular structures are often an indication of an ongoing merger.
Looked at individually, very large cooling flows are almost never
associated with very irregular or bimodal clusters,
which are likely merger candidates
(\nolinebreak\cite{H88};
\cite{Eea92}).
There are some cases of moderate cooling flows in merging clusters;
in most cases, these appear to be early-stage mergers where the merger
shocks haven't yet reached the cooling core of the cluster.
Examples may include Cygnus-A
(\nolinebreak\cite{Aea84};
\cite{OLMH};
Markevitch et al.\ 1999) and
Abell~85
(\nolinebreak\cite{KSR}).
There also are a large number of merging clusters at a more advanced stage
with relatively small cooling cores,
both in terms of the cooling rate and the physical radius;
Abell~2065
(Markevitch et al. 1999)
may be an example.
Recently, Chandra Observatory X-ray images have shown a number of merging
clusters with rapidly moving cores of cool gas
(the ``cold fronts'' discussed below in \S~\ref{sec:cf_coldfront}).
In these systems, the cooling flows appear to have survived, at least to
the present epoch in the merger.

It is unclear exactly how and under what circumstances mergers
disrupt cooling flows.
The cooling flows might be disrupted
by tidal effects,
by shock heating the cooler gas,
by removing it dynamically from the center of the cluster due to ram pressure,
by mixing it with hotter intracluster gas,
or by some other mechanism.
Numerical hydrodynamical simulations are needed to study the mechanisms
by which cooling flows are disrupted.
This is a relatively unexplored area, largely because the small spatial
scales and rapid cooling time scales in the inner regions of cooling flows
are still a significant challenge to the numerical resolution of hydrodynamical
codes.
McGlynn \& Fabian (1984)
\nocite{MF}
argued that mergers disrupted cooling flows, but this was based on
purely N-body simulations.
Recently,
G\`omez et al.\ (2001)
\nocite{gomez}
have made hydrodynamical simulations of the effects of head-on mergers
with relatively small subclusters (1/4 or 1/16 of the mass of the main
cluster) on a cooling flow in the main cluster.
They find that the mergers disrupt the cooling flow in some cases, but
not in others.
Their simulations suggest that the disruption is not due to
tidal or other gravitational effects.

Another possibility is that the merger shocks heat up the cooling flow gas
and stop the cooling flow.
In the simulations, this does not appear to be the main mechanism of
cooling flow disruption.
There are a number of simple arguments which suggest that merger shocks
should be relatively inefficient at disrupting cooling flows.
First, it is difficult for these shocks to penetrate the high densities
and steep density gradients associated with cooling flows,
and the merger shocks would be expected to weaken as they climb these steep
density gradients.
Even without this weakening, merger shocks have low Mach numbers, and only
produce rather modest increases in temperature ($\la$ a factor of 2).
These small temperature increases are accompanied by significant compressions.
As a result, shock heating actually decreases the cooling time due to
thermal bremsstrahlung emission for shocks with Mach numbers
${\cal M} \le ( 21 + 12 \sqrt{3} )^{1/2} \approx 6.5$.
It is likely that the shocked gas will eventually expand, and adiabatic
expansion will lengthen the cooling time.
However, even if the gas expands to its preshock pressure, the increase
in the cooling time is not very large.
For a ${\cal M} = 2$ shock, the final cooling time after adiabatic
expansion to the original pressure is only about 18\% longer than the
initial cooling time.

The simulations by G\`omez et al.\ suggest that the main mechanism for
disrupting cooling flows is associated with the ram pressure of gas from
the merging subcluster.
The gas in the cooling flow is displaced, and may eventually mix with
the hotter gas
(see also \cite{RiS}).
Earlier, Fabian \& Daines (1991)
\nocite{FD}
had argued that ram pressure, rather than shock heating, was the
main mechanism for disrupting cooling flows.
Assuming this is the case, one expects that the merger will remove
the cooling flow gas at radii which satisfy
\begin{equation} \label{eq:ram}
\rho_{\rm sc} v_{\rm rel}^2 \ga P_{\rm CF} ( r ) \, ,
\end{equation}
where $P_{\rm CF} ( r )$ is the pressure profile in the cooling flow,
$\rho_{\rm sc}$ is the density of the merging subcluster gas at
the location of the cooling flow, and $v_{\rm rel}$ is the relative
velocity of the merging subcluster gas and the cooling flow.
G\`omez et al.\ (2001) find that this relation provides a reasonable
approximation to the disruption in their simulations.

The pressure profile in the cooling flow gas prior to the merger is
determined by the condition of hydrostatic equilibrium.
If the cluster gravitational potential has a wide core within which
the potential is nearly constant
(e.g., as in a King model),
then the cooling flow pressure will not increase rapidly into the center.
In this case, once the merger reaches the central regions of the cluster,
if the ram pressure is sufficient to remove the outer parts of the 
cooling flow, it should be sufficient to remove nearly all of the cooling
flow.
On the other hand, if the cluster potential is sharply peaked
(as in a NFW profile, \cite{NFW}), the merger may remove the
outer parts of the cooling flow but not the innermost regions.
Thus, the survival and size of cool cores in merging clusters
can provide evidence on whether clusters have sharply peaked
potentials.
Markevitch et al.\ (1999) applied this argument to the two small
cool cores in the merging cluster Abell~2065, and concluded that
steep central potentials, consistent with the NFW model, were needed.

\subsection{Cold Fronts}\label{sec:cf_coldfront}

One of the more dramatic early discoveries with the Chandra X-ray
Observatory was the presence of very sharp surface brightness
discontinuities in merging clusters of galaxies
(e.g., \cite{forman}).
A pair of such discontinuities were first seen in the public science
verification data on the Abell~2142 cluster
(\nolinebreak\cite{Mea2000}).
Initially, it seemed likely that these were merger shocks.
However, temperature measurements showed that this was not the
case.
The high X-ray surface brightness regions were both dense and {\it cool};
thus, the gas in these regions had a lower specific entropy than the
gas in the less dense regions.
Shocks are irreversible processes which must increase the entropy as
the gas is compressed.
Also, the gas pressure appeared to be continuous across the density
discontinuity.
The lack of a pressure jump and the incorrect sign of the temperature
and entropy variations showed that these features could not be shocks
(\nolinebreak\cite{Mea2000}).

Instead, they appear to be contact discontinuities between hot, diffuse gas
and a cloud of colder, denser gas
(\nolinebreak\cite{Mea2000}).
The cold cloud is moving rapidly through the hotter gas;
Vikhlinin et al.\ (2001b) refer to this situation as a ``cold front.''
\nocite{VMMb}
The cold cloud is being distorted and, presumably, stripped by the hot
gas, but has survived to the epoch of the observation.
Markevitch et al.\ (2000) argue that the source of the cold cloud is the
core of one of the merging subclusters;
in Abell~2142, both of the subcluster cores appear to have survived.
There are several reasons why the core gas is much denser than the
surrounding hot gas.
First, prior to mergers, clusters are generally stably-stratified,
with the denser, lower entropy gas at the center.
Thus, the core density will be much higher than the density of the outer gas
in a cluster.
Second, in hierarchical large scale structure, smaller subclusters generally
form from denser perturbations, so small merging subclusters may have cores
with quite high densities.
Third, many rich and poor clusters have cooling flows at their centers
(\S~\ref{sec:cf_cflow}), and these regions have very high densities and
relatively low temperatures.
As noted above, cooling flows do appear to be able to partially survive in
mergers, at least for some period.

Subsequently, cold fronts have been observed in a number of other clusters,
including
Abell~3667
(\nolinebreak\cite{VMMb}),
RXJ1720.1+2638
(\nolinebreak\cite{Maz01}),
Abell~85
(\nolinebreak\cite{KSR}),
and possibly
Abell~754 and Abell~2163
(\nolinebreak\cite{MVMV}).
The most detailed analysis has been made for Abell~3667,
and my discussion closely follows the arguments given in
Vikhlinin et al.\ (2001b).
\nocite{VMMb}

\subsubsection{Kinematics of Cold Fronts}\label{sec:cf_coldfront_kin}

\begin{figure}[t]
\vskip2.2truein
% \vskip3.8truein
% \special{psfile=coldfront.ps hscale=50 vscale=50
\includegraphics{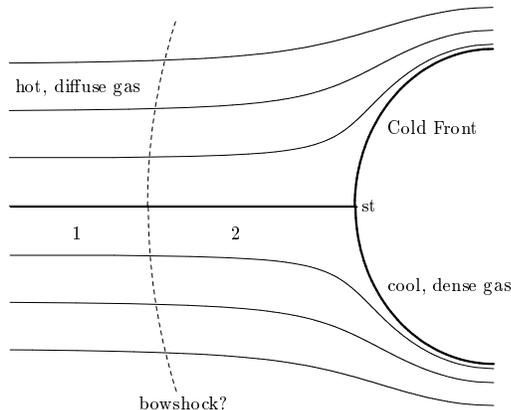}
\caption[]{A schematic diagram of flow around a ``cold front'' in a cluster
merger.
The heavy solid arc at the right represents the contact discontinuity
between the cold, dense cold core gas, and the hotter, more diffuse gas
from the outer regions of the other cluster.
The cold core is moving toward the left relative to the hotter gas.
The narrow solid lines are streamlines of the flow of the hotter gas
around the cold core.
The region labelled ``1'' represent the upstream, undisturbed hot gas.
If the cold front is moving transonically (${\cal M}_1 > 1$), then
the cold front will be preceded by a bow shock, which is shown as a
dashed arc.
The stagnation point, where the relative velocity of the cooler dense
gas and hotter diffuse gas is zero, is marked ``st''.
\label{fig:cold_front}}
\end{figure}

As discussed extensively in Vikhlinin et al.\ (2001b),
\nocite{VMMb}
the variation in the density, pressure, and temperature of the gas
in a cold front can be used to determine the relative velocity of
cold core.
This technique is analogous to that for merger shocks
discussed above
(eqs.~\ref{eq:jumpv} \& \ref{eq:jumpC}).
The geometry is illustrated in Figure~\ref{fig:cold_front},
which is drawn in the rest frame of the cold core.
We assume that the cold core has a smoothly curved, blunt front edge.
The normal component of the flow of hot gas past the surface of the
cold core will be zero.
There will be at least one point where the flow is perpendicular to
the surface of the cold core, and the flow velocity of the hot gas will
be zero at this stagnation point (``st'' in Fig.~\ref{fig:cold_front}).
Far upstream, the flow of the hot gas will be undisturbed at the velocity
of the cold core relative to the hotter gas, $v_1$.
Let $c_{s1}$ be the sound speed in this upstream gas, and
${\cal M}_1 \equiv v_1 / c_{s1} $ be the Mach number of the motion
of the cold core into the upstream gas.
If ${\cal M}_1 > 1$, a bow shock will be located ahead of the cold front.

The ratio of the pressure at the stagnation point to that far upstream
is given by
(e.g., \cite{LanLif}, \S~114).
% (e.g., \cite{LanLif}, \S~122).
\begin{equation} \label{eq:pst}
\frac{P_{\rm st}}{P_1} = \left\{
\begin{array}{cl}
\left( 1 + \frac{\gamma_{\rm ad} - 1}{2} {\cal M}_1^2
\right)^{\frac{\gamma_{\rm ad}}{\gamma_{\rm ad} - 1}} \, , & 
{\cal M}_1 \le 1 \, , \\
{\cal M}_1^2 \,
\left( \frac{\gamma_{\rm ad} + 1}{2}
\right)^{\frac{\gamma_{\rm ad} + 1}{\gamma_{\rm ad} - 1}} \,
\left( \gamma_{\rm ad} - \frac{\gamma_{\rm ad} - 1}{2 {\cal M}_1^2} \,
\right)^{- \frac{1}{\gamma_{\rm ad} - 1}} \, , &
{\cal M}_1 > 1 \, . \\
\end{array}
\right.
\end{equation}
The ratio $( P_{\rm st} / P_1 )$ increases continuously and monotonically
with ${\cal M}_1$.
Thus, 
in principle, measurements of $P_1$ and $P_{\rm st}$ in the hot gas
could be used to determine ${\cal M}_1$.
The pressures would be determined from X-ray spectra and images.
In practice, the emissivity of the hot gas near the stagnation point
is likely to be small.
However, the pressure is continuous across the cold front, so the
stagnation pressure can be determined just inside of the cold core,
where the X-ray emissivity is likely to be much higher.
Once ${\cal M}_1$ has been determined, the velocity of the encounter is
given by
$v_1 = {\cal M}_1 c_{s1}$.

If the motion of the cold core is transonic (${\cal M}_1 > 1$),
one can also determine the velocity from the temperature and/or density
jump at the bow shock
(eqs.~\ref{eq:jumpv} \& \ref{eq:jumpC}).
If the bow shock can be traced to a large transverse distance and forms
a cone, the opening angle of this Mach cone corresponds to the Mach
angle, $\theta_M \equiv \csc^{-1} ( {\cal M}_1 )$.
However, variations in the cluster gas temperature may lead to distortions
in this shape.

\begin{figure}[t]
\vskip2.8truein
% \special{psfile=stand_off.ps hscale=40 vscale=40 angle=-90
\includegraphics{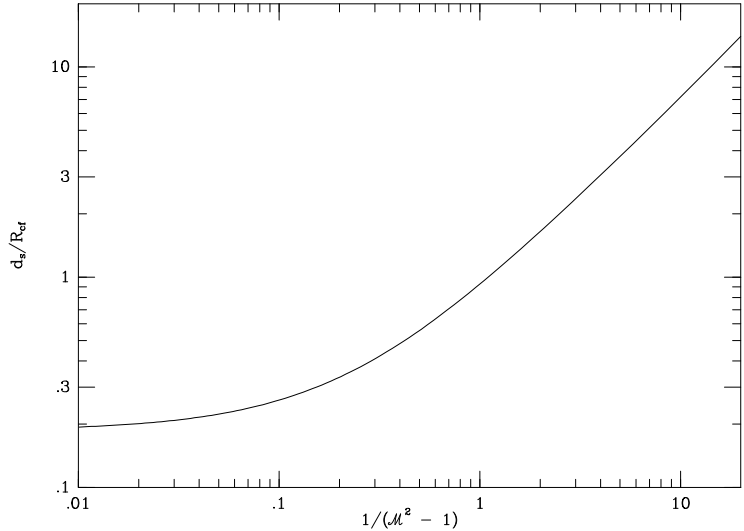}
\caption[]{The ratio of the stand-off distance of the bow shock $d_s$ to
the radius of curvature $R_{\rm cf}$ of the stagnation region of the cold
front, as a function of $1 / ( {\cal M}_1^2 - 1 )$, where
${\cal M}_1$ is the Mach number.
This is for a spherical cold front and $\gamma_{\rm ad} = 5/3$.
% The solid line assumes that the cold front is a sphere, while the dashed
% line is for a paraboloid shape
% (\nolinebreak\cite{rusanov}).
\label{fig:stand_off}}
\end{figure}

The distance between the stagnation point and the closest point on
the bow shock (the shock ``stand-off'' distance $d_s$) can also be used
to estimate the Mach number of the motion of the cold front
% (Vikhlinin et al.\ 2001b).
(\nolinebreak\cite{VMMb}).
The ratio of $d_s$ to the radius of curvature of the cold front
$R_{\rm cf}$ depends on the Mach number ${\cal M}_1$
and on the shape of the cold front.
Figure~\ref{fig:stand_off} shows the values of $d_s / R_{\rm cf}$ as a
function of $( {\cal M}_1^2 - 1 )^{-1}$ for
a spherical cold front
(\nolinebreak\cite{schreier}).
% either a spherical or paraboloid surface
% (e.g., \cite{rusanov}).
Although there is no simple analytic expression for the stand-off distance
which applies to all shapes of objects,
a fairly general approximate method to calculate $d_s$ has been given
by
Moekel (1949),
\nocite{moekel}
and some simple approximate expressions exist for a number of simple
geometries
(\nolinebreak\cite{guy};
\cite{radvogin}).
The stand-off distance increases as the Mach number approaches unity;
thus, this method is, in some ways, a very sensitive diagnostic for
the Mach number for the low values expected in cluster mergers.
On the other hand, the stand-off distance also depends strongly on the
shape of the cold front as the Mach number decreases.
The application of this diagnostic to observed clusters is strongly
affected by projection effects.
Because the radius of curvature of the bow shock is usually greater
than that of the cold front
(\nolinebreak\cite{rusanov}),
projection effects will generally cause $d_s$ to be overestimated and
${\cal M}_1$ to be underestimated.
Projection effects also make the true shape of the cold front uncertain.

These techniques have been used to determine the merger velocities
from cold fronts in
Abell~3667
(\nolinebreak\cite{VMMb}),
RXJ1720.1+2638
(\nolinebreak\cite{Maz01}),
and
Abell~85
(\nolinebreak\cite{KSR}).

\subsubsection{Width of Cold Fronts}\label{sec:cf_coldfront_nar}

One remarkable aspect of the cold fronts observed with the Chandra
Observatory in several clusters is their sharpness.
In Abell~3667, the temperature changes by about a factor of two
across the cold front (\nolinebreak\cite{VMMb}), and the accompanying
change in the X-ray surface brightness occurs
in a region which is narrower than 2 kpc
(\nolinebreak\cite{VMMb}).
This is less than the mean-free-path of electrons in this region.
The existence of this very steep temperature gradient and similar results
in other merging clusters with cold fronts requires
that thermal conduction be suppressed by a large factor
(\nolinebreak\cite{EF};
Vikhlinin, Markevitch, \& Murray 2001a,b)
\nocite{VMMa}
\nocite{VMMb}
relative to the classical value in an unmagnetized plasma
(e.g., \cite{spitzer}).
It is likely that this suppression is due to the effects of the
intracluster magnetic field.
It is uncertain at this point whether this is due to a generally tangled
magnetic field (in which case, heat conduction might be suppressed
throughout clusters), or due to a tangential magnetic field specific
to the tangential flow at the cold front
(\nolinebreak\cite{VMMa}).

Because of the tangential shear flow at the cold front
(Fig.~\ref{fig:cold_front}),
the front should be disturbed and broadened by the Kelvin-Helmholtz
(K-H)
instability.
Vikhlinin et al.\ (2001a) argue that the instability is suppressed by
\nocite{VMMa}
a tangential magnetic field, which is itself generated by the
tangential flow.
This suppression requires that the magnetic pressure $P_B$ be a non-trivial
fraction of the gas pressure $P$ in this regions, $P_B \ga 0.1 P$.
The required magnetic field strength in Abell~3667 is $B \sim 10$ $\mu$G.

\section{Nonthermal Physics of Merger Shocks} \label{sec:nonthermal}

Cluster mergers involve shocks with velocities of $\sim$2000 km/s.
Radio observations of supernova remnants indicate that shocks with
these velocities can accelerate or reaccelerate relativistic electrons
and ions
(e.g., \cite{BE}).
In order to explain the general radio emission of supernova remnants, one
requires that shocks in these systems generally convert a few percent
of the shock energy into relativistic electrons.
Even more energy may go into relativistic ions.
Thus, one might expect that the intracluster medium would contain relativistic
particles or cosmic rays, in addition to the hot thermal gas so evident
in X-ray images.
Given that all of the thermal energy content of the intracluster gas in
clusters is due to shocks with velocities of $\ga$$10^3$ km/s,
it seems likely that relativistic electrons and ions will have been
accelerated with a total energy content of a few percent of the thermal
energy in the hot gas.
In massive, X-ray luminous clusters, the total thermal energy content in
the ICM is $\ga$$3 \times 10^{63}$ ergs.
Thus, merger or accretion shocks may have accelerated cosmic ray particles
with a total energy content of $\ga$$10^{62}$ ergs.
This would make clusters the largest individual sources of relativistic
particles in the Universe;
this energy probably exceeds that produced in active galactic nuclei,
such as quasars and radio galaxies.

In a major merger, the thermal energy content of a cluster can be significantly
increased by the merger shocks (\S~\ref{sec:thermal}).
Thus, shock acceleration or reacceleration processes in a single merger
may produce cosmic ray particles with a total energy of
$\sim$$10^{62}$ ergs.
Thus, one would expect significant nonthermal effects associated with
cluster mergers.

\begin{figure}[t]
\vskip2.8truein
% \special{psfile=losses.ps hscale=40 vscale=40 angle=-90
\includegraphics{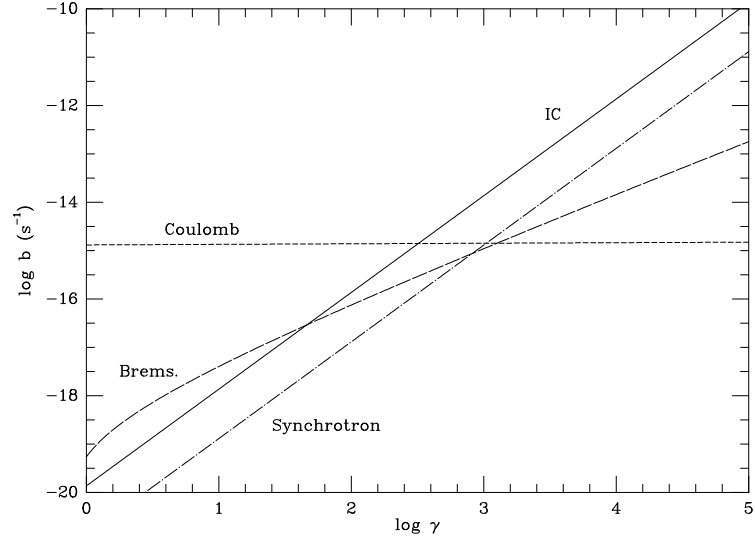}
\caption[]{Values of the electron loss functions $b ( \gamma )$ for 
inverse Compton (IC) emission,
Coulomb losses,
synchrotron emission, and
bremsstrahlung emission as a function of $\gamma = E / ( m_e c^2 )$.
The values assume $n_e = 10^{-3}$ cm$^{-3}$, $B = 1 \, \mu$G, and
redshift $z = 0$.
\label{fig:losses}}
\end{figure}

\subsection{Particle Lifetimes and Losses} \label{sec:nonthermal_life}

Clusters are also very good storage locations for cosmic rays.
These particles gyrate around magnetic field lines in the ICM.
The magnetic field is frozen-in to the ionized thermal ICM, which is,
in turn, bound by the gravitational field of the cluster.
Thus, the relativistic particles cannot simply stream out of a cluster.
They can diffuse out along magnetic field lines.
Diffusion is limited by scattering off of fluctuations in the magnetic
field, and the rate is uncertain.
However, under reasonable assumptions, the diffusion coefficient is
approximately
(Berezinsky, Blasi, \& Ptuskin 1997;
\nocite{BBP}
% (\nolinebreak\cite{BBP};
\cite{CB})
\begin{equation} \label{eq:diffusion}
D( E ) \approx 2 \times 10^{29} \,
\left( \frac{E}{1 \, {\rm GeV}} \right)^{1/3} \,
\left( \frac{B}{1 \, \mu{\rm G}} \right)^{-1/3} \,
{\rm cm}^{2} \, {\rm s}^{-1} \, ,
\end{equation}
where $E$ is the particle energy and $B$ is the ICM magnetic field.
The average time scale to diffuse out to a radius of $R$ is about
(\nolinebreak\cite{BBP}; \cite{CB})
\begin{equation} \label{eq:diff_time}
t_{\rm diff} \approx \frac{R^2}{6 D(E)}
\approx 
1 \times 10^{12}
\left( \frac{R}{2 \, {\rm Mpc}} \right)^{2} \,
\left( \frac{E}{1 \, {\rm GeV}} \right)^{-1/3} \,
\left( \frac{B}{1 \, \mu{\rm G}} \right)^{1/3} \, {\rm yr} \, .
\end{equation}
Thus, under reasonable assumptions for the diffusion coefficient, particles
with energies $\la$$10^6$ GeV have diffusion times which
are longer than the Hubble time.

\begin{figure}[t]
\vskip2.8truein
\includegraphics{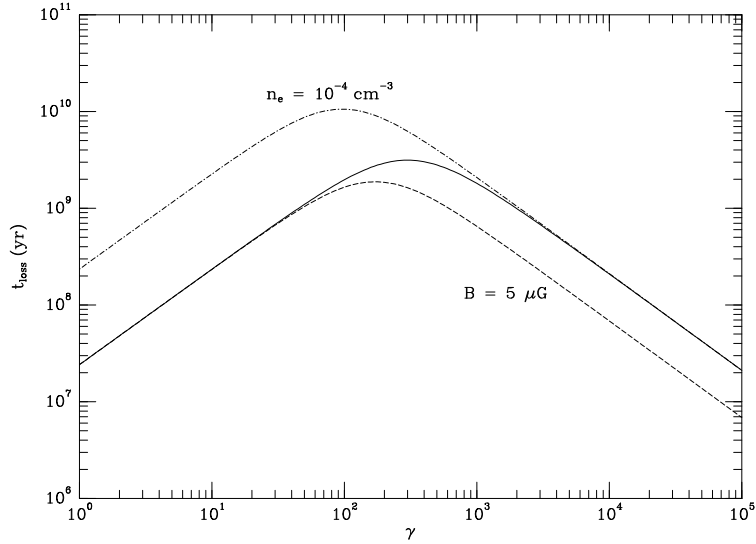}
\caption[]{
The solid curve gives the instantaneous loss time scale $t_{\rm loss}$
for relativistic electrons in a cluster with an electron density
of $n_e = 10^{-3}$ cm$^{-3}$ and a magnetic field of $B = 1$ $\mu$G.
The short-dash curve is for $B = 5$ $\mu$G, while the dash-dot curve
is for $n_e = 10^{-4}$ cm$^{-3}$.
\label{fig:lifetime}}
\end{figure}

Relativistic particles can lose energy, and this can effectively remove
them from the cosmic ray population.
The time scales for energy loss by ions are generally longer than
the Hubble time.
Electrons suffer losses due to interactions
with ambient radiation fields (via inverse Compton [IC] emission),
with the cluster magnetic field (via synchrotron emission), and
with the intracluster gas (via Coulomb interactions and bremsstrahlung
emission).
However, the ICM is an extremely diffuse medium, and these losses
are relatively small, at least as compared to the interstellar gas in our
Galaxy.
The gas density is low ($n_e \sim 10^{-3}$ cm$^{-3}$), reducing Coulomb
and bremsstrahlung losses.
The radiation fields are dilute, with the Cosmic Microwave Background (CMB)
radiation providing the majority of the energy density.
Magnetic fields are relatively weak;
if the cluster fields are mainly smaller than 3 $\mu$G, then synchrotron
losses are smaller than IC losses.

Let the energy of an electron be $E \equiv \gamma m_e c^2$, where
$\gamma$ is the Lorentz factor.
Then, the energy loss of an electron can be written as
\begin{equation} \label{eq:loss}
\frac{d \gamma}{dt} = \frac{1}{m_e c^2} \, \frac{d E}{dt} = - b( \gamma ) \, ,
\end{equation}
assuming the loss is continuous.
The values of the loss functions $b ( \gamma )$ for various processes are
shown in Figure~\ref{fig:losses}
(\nolinebreak\cite{S1}).
It is clear that IC and synchrotron losses are dominant at high energies
($\gamma \ga 200$ or $E \ga 100$ MeV),
while Coulomb losses dominate at low energies
($\gamma \la 200$ or $E \la 100$ MeV).

One can define an instantaneous time scale for energy losses as
$ t_{\rm loss} \equiv \gamma / b( \gamma ) = E / ( dE/dt )$.
Values for this loss time scale at the present epoch ($z = 0$) are
shown in Figure~\ref{fig:lifetime}
(\nolinebreak\cite{S1}).
The solid curve gives values assuming an average electron density of
$n_e = 10^{-3}$ cm$^{-3}$ and a magnetic field of $B = 1$ $\mu$G.
For values of the magnetic field this small or lower, synchrotron losses
are not very significant, and $t_{\rm loss}$ is nearly independent of $B$.
The short dashed curve shows the effect of increasing the magnetic
field to $B = 5$ $\mu$G;
the losses at high energies are increased, and the loss time scales
shortened.
The dash--dot curve shows the loss time scale if the electron density is
lowered to $n_e = 10^{-4}$ cm$^{-3}$.
This reduces the losses at low energies, and increases the loss times
there.
Although high energy electrons lose energy rapidly due to IC and
synchrotron emission, electrons with Lorentz factors of $\gamma \sim 300$
(energies $\sim 150$ MeV)
have long lifetimes of $\sim 3 - 10$ Gyr, which are comparable to the
likely ages of clusters
(\nolinebreak\cite{SL};
\cite{S1}).
Thus, clusters of galaxies can retain low energy electrons
($\gamma \sim 300$) and nearly all cosmic ray ions for a significant
fraction of a Hubble time.

\subsection{Sources of Relativistic Particles} \label{sec:nonthermal_src}

What are the sources for relativistic particles in clusters?
One possibility is that these particles come from active galaxies
(quasars, radio galaxies, etc.; e.g.,
\cite{BC}).
Because luminous active galaxies were more common in the past,
most of the cosmic ray particles would probably have been formed in the
past.
Another possibility is that these particles were generated as part of
star formation in normal galaxies,
either at the sites of star formation and supernova, or in galactic winds
(e.g., \cite{AV}).
The galaxies in the inner regions of clusters today are mainly
elliptical and S0 galaxies, which have old stellar populations.
Thus, most of their star formation, and most of the particle production
associated with it, probably occurred in the distant past.
In any case, if AGNs or star bursts produced most of particles in clusters
directly, then the cosmic ray populations in clusters would have no clear
relation to mergers.
I concentrate here on models in which the particles were either produced
directly in mergers, or are the secondary products of particles produced
in mergers, and/or were reaccelerated in mergers.

\subsubsection{Particle Acceleration in Shocks}
\label{sec:nonthermal_src_acceler}

Radio observations of supernova remnants indicate that
shocks with $v \ga 10^3$ km/s convert at least a few
percent of the shock energy into the acceleration of relativistic
electrons
(e.g., \cite{BE}).
Even more energy may go into relativistic ions.
Thus, merger shocks seem like a natural acceleration site for relativistic
particles.
It is worth noting that there are significant differences between merger
shocks and those associated with supernova blast waves.
The merger shocks have relatively small Mach numbers, and as a result
have smaller compressions.
The ICM which enters the merger shock is hot.
This means that the shocks are subsonic in the electrons;
the preshock electrons have thermal velocities which are much greater than the
shock velocities.
On the other hand, the Alfv\'en Mach numbers (${\cal M}_A \equiv v_s / v_A$
where $v_A = B^2 / ( 4 \pi \rho )$ is the Alfv\'en speed) for merger
shocks can be quite large, ${\cal M}_A \ga 30$.
For some aspects of shock acceleration, the Alfv\'en Mach number is more
relevant than the hydrodynamical Mach number.

Assuming that particles scatter repeatedly across the shock,
these particles will undergo first-order Fermi shock acceleration.
If the accelerating particles are treated as test particles, kinetic
theory indicates that the particle spectrum is a power-law in the momentum
$p$
(\nolinebreak\cite{bell};
\cite{BO}):
\begin{equation} \label{eq:powerlaw}
N ( p ) \, dp  =  N_o
\left( \frac{p}{mc} \right)^{- \mu} \, \frac{dp}{mc} \, ,
\qquad p_l \le p \le p_u
\end{equation}
where $m$ is the particle mass.
Here, $N(p) \, dp$ is the number of particles with momenta between
$p$ and $p + dp$,
and $p_l$ ($p_u$) are the lower (upper) limits on the particle spectrum.
If the particles are accelerated from nearly thermal energies,
then the lower limit may be associated with the production of a
nonthermal tail at the high energy end of the ICM thermal particle
distribution.
The upper limit may correspond to the highest energy for which
acceleration is efficient
(e.g., \S~\ref{sec:nonthermal_emit_cr}).
The particle spectrum expressed in terms of the Lorentz factor is
\begin{eqnarray}
N( \gamma ) \, d \gamma & = & N_o 
\left( \gamma^2 - 1 \right)^{-( \mu + 1 ) / 2} \, \gamma \, d \gamma
\, ,\nonumber \\
& \approx & N_o \gamma^{- \mu } \, d \gamma \, , \qquad \gamma \gg 1 \, .
\label{eq:espectrum}
\end{eqnarray}
The energy spectrum is given by $N(E) \, dE = N( \gamma ) \, d \gamma$,
with $E = \gamma m c^2$.
Thus, the energy spectrum for the shock acceleration of relativistic particles
is also expected to be a power-law.

For shock acceleration, the exponent is
\begin{equation} \label{eq:power}
\mu = \frac{C + 2}{C - 1} \, ,
\end{equation}
where $C$ is the shock compression
(eq.~\ref{eq:jumpM} \& \ref{eq:jumpC}).
Strong shocks give $C = 4$ and $\mu = 2$, which is in reasonable agreement
with the radio observations of supernova remnants.
Merger shocks have $C \approx 2 - 3$, which leads to $\mu \approx 4 - 2.5$.
Thus, the particle spectra produced by merger shocks are expected to be
significantly steeper than those generated by supernova remnant blast
waves.

\subsubsection{Reacceleration by Merger Shocks}
\label{sec:nonthermal_src_reacceler}

Merger shocks may reaccelerate pre-existing relativistic particles, rather
than produce new particles from the thermal ICM.
This mechanism has been proposed to explain the radio halo in the Coma
cluster and other halos
(\nolinebreak\cite{BSFGa},b).
\nocite{BSFGa}
In this model, the reacceleration occurs gradually over an extended period
of time.

Radio relics might also be due to the reacceleration of relatistic
particles injected as some time in the past by radio galaxies
(\nolinebreak\cite{EnsslinB}).
In this case, one would only expect to see relics associated with a small
fraction of merger shocks;
one would require both a merger shock and a pre-existing radio population.
If the old radio plasma continues to be separated from the thermal plasma
(a radio ``ghost,'' \cite{ensslin}), then the merger shock will be subsonic
in the relativistic radio plasma.
Thus, rather than reacceleration, the merger shock might re-energize the
radio plasma by adiabatic compression.

\subsubsection{Turbulent Acceleration Following a Merger}
\label{sec:nonthermal_src_turb}

Cluster mergers may produce a significant level of turbulence in
the ICM,
and this could lead to turbulent acceleration or reacceleration of
relativistic electrons
(\nolinebreak\cite{EW}).
This is second order Fermi acceleration.
Turbulent reacceleration has been suggested to explain radio
halos in clusters
(\nolinebreak\cite{BSFGa},b).
\nocite{BSFGa}
Radio halos have only been found in merging clusters.
However, their smooth distributions and central locations
suggest that they are not confined to the region currently passing through
a merger shock.
Turbulent acceleration following the passage of merger shocks
might explain these properties.

\subsubsection{Secondary Electron Production}
\label{sec:nonthermal_src_second}

Another source of relativistic electrons is
the decay of charged mesons generated in cosmic ray
ion collisions
(\nolinebreak\cite{dennison};
\cite{vestrand};
\cite{CB}).
The reactions involved are
\begin{eqnarray}
p + p & \rightarrow & \pi^{\pm} + X \, , \nonumber \\
\pi^{\pm} & \rightarrow & \mu^{\pm} + \nu_\mu \, ( \bar{\nu}_\mu ) \, \,
\nonumber \\
\mu^{\pm} & \rightarrow & e^{\pm} + \bar{\nu}_\mu \, ( {\nu}_\mu )
+ \nu_e \, ( \bar{\nu}_e ) \, .
\label{eq:secondary}
\end{eqnarray}
Here, $X$ represents some combination of protons, neutrons, and/or other
particles.
The electrons (and positrons) produced by this mechanism are referred to as
secondary electrons.
If the primary cosmic ray ions are due to AGNs or star bursts, this
process might have no connection with cluster mergers.
On the other hand, the ions might have been accelerated or reaccelerated
by cluster merger shocks or turbulence associated with cluster mergers.
Recently, 
En{\ss}lin (2001)
\nocite{ensslin}
proposed that primary ions in clusters were originally produced by AGN or
starbursts, but had lost most of their energy due to adiabatic losses.
He argued that these ions are reaccelerated by merger shocks, and
the subsequent secondary electrons make cluster radio halos.

\subsection{Models for Merger Shocks and Primary Electrons}
\label{sec:nonthermal_model}

Here, I describe the results of some models for the population of
relativistic electrons in clusters, assuming they are primary electrons
accelerated in merger shocks
(\nolinebreak\cite{S1}; \S~\ref{sec:nonthermal_src_acceler}).
The populations of cosmic ray electrons in clusters depends on their merger
histories.
Because low energy electrons have long lifetimes, one expects to find
a large population of them in most clusters (any cluster which has had
a significant merger since $z \sim 1$).
On the other hand, higher energy electrons ($E \ga 1$ GeV) have
short lifetimes (shorter than the time for a merger shock to cross
a cluster).
Thus, one only expects to find large numbers of higher energy primary
electrons in clusters which are having or have just had a merger.
These conclusions follow from a large number of detailed models
of the evolution of the integrated electron population in clusters
(\nolinebreak\cite{S1}).
Two recent cluster merger simulations have included particle acceleration
approximately
(\nolinebreak\cite{RBS}; \cite{TN}), and they reach similar conclusions.

\begin{figure}[t]
\vskip2.8truein
% \special{psfile=espect.ps hscale=40 vscale=40 angle=-90
\includegraphics{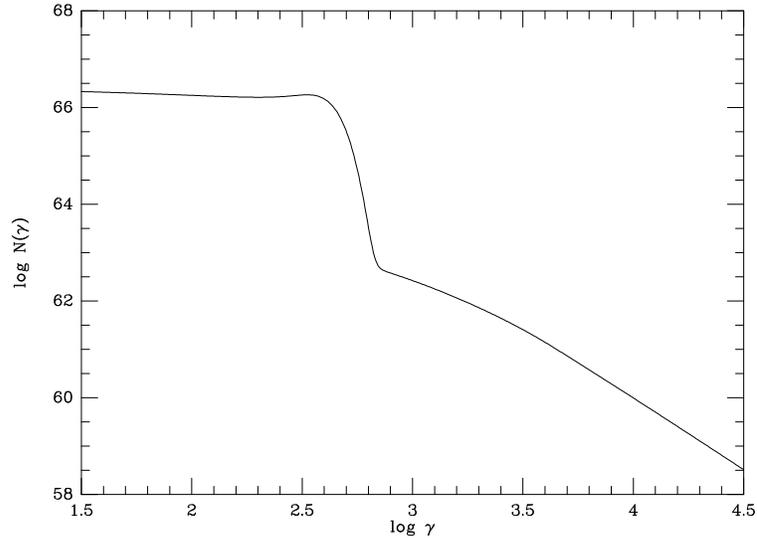}
\caption[]{
A typical model for the relativistic electron population in a cluster of
galaxies.
The lower energy electrons are due to all of the mergers in the cluster
history, while the high energy electrons are due to a small current
merger.
\label{fig:espect}}
\end{figure}

% \begin{figure}[t]
% \special{psfile=espect.ps angle=-90 hscale=32 vscale=32
% 	voffset=+12 hoffset=-17}
% \special{psfile=ic.ps angle=-90 hscale=32 vscale=32
% 	voffset=+12 hoffset=213}
% \centerline{\null}
% \centerline{\null}
% \noindent\hskip2.66truein (a) \hskip2.94truein (b) \hfill
% \vskip1.70in
% \caption[]{(a) A typical model for the relativistic electron population
% in a cluster of galaxies.
% The lower energy electrons are due to all of the mergers in the cluster
% history, while the high energy electrons are due to a small current
% merger.
% (b)
% The IC spectrum from the same model (solid curve).
% The dashed curve is a 7 keV thermal bremsstrahlung spectrum.\hfill
% \label{fig:espect}
% }
% \end{figure}

% Fig.~\ref{fig:espect}(a) shows the electron spectrum in a cluster
Figure~\ref{fig:espect} shows the electron spectrum in a cluster
with a typical history.
Most of the electron energy is in electrons with $\gamma \sim 300$, which
have the longest lifetimes.
These electrons are produced by mergers over the entire history
of the cluster.
This cluster also has a small ongoing merger which produces
the high energy tail on the electron distribution.
In cluster models without a current merger, the high energy tail would be
missing.

Most of the emission from these electrons is due to IC, and the
% resulting spectrum is shown in Fig.~\ref{fig:espect}(b).
resulting spectrum is shown in Figure~\ref{fig:ic}.
For comparison, thermal bremsstrahlung with a typical rich cluster
temperature and luminosity is shown as a dashed curve.
Figure~\ref{fig:ic} shows that clusters should be strong
sources of extreme ultraviolet (EUV) radiation.
Since this emission is due to electrons with $\gamma \sim 300$ which
have very long lifetimes, EUV radiation should be a common feature
of clusters
(\nolinebreak\cite{SL}).

In clusters with an ongoing merger, the higher energy electrons will
produce a hard X-ray tail via IC scattering of the Cosmic Microwave
Background (CMB);
the same electrons will produce diffuse radio synchrotron emission.

\subsection{Nonthermal Emission and Mergers}
\label{sec:nonthermal_emit}

\subsubsection{Radio Halos and Relics } \label{sec:nonthermal_emit_radio}

The oldest and most detailed evidence for nonthermal populations in
clusters comes from the radio.
A number of clusters of galaxies are known to contain large-scale
diffuse radio sources which have no obvious connection to individual
galaxies in the cluster
(\nolinebreak\cite{Gea}).
These sources are referred to as radio halos when they appear projected
on the center of the cluster, and are called relics when they are found
on the cluster periphery (although they have other distinctive properties).
In all cases of which I am aware, they have been found in clusters which
show significant evidence for an ongoing merger
(\nolinebreak\cite{Gea};
\cite{Fer1};
\cite{Fer2}).
Since these source are discussed extensively in another chapter of this
book
(\nolinebreak\cite{GF}), I won't discuss them in any more detail here.

% ?? give analytic arguments on luminosities.

\subsubsection{EUV/Soft X-ray Emission} \label{sec:nonthermal_emit_euv}

Excess EUV emission has apparently been detected with the EUVE satellite in
six clusters
(Virgo, Coma, Abell 1795, Abell 2199, Abell 4038, \& Abell 4059;
\cite{Lea1},b;
\cite{BB};
\cite{MLL};
Bowyer, Bergh\"ofer, \& Korpela 1999;
\cite{Kea};
\cite{Lea3},b;
Bergh\"ofer, Bowyer, \& Korpela 2000a,b;
Bonamente, Lieu, \& Mittaz 2000).
\nocite{BBK1}
\nocite{BBK2}
\nocite{BBK3}
\nocite{BLM}
\nocite{Lea2}
\nocite{Lea4}
In fact, the EUVE satellite appears to have detected all of the clusters
it observed which are nearby, which have long integration times, and 
which lie in directions of low Galactic column where detection is
possible at these energies.
However, the EUV detections and claimed properties of the clusters
remain quite controversial
(\nolinebreak\cite{BB};
\cite{AB};
\cite{BBK2};
\cite{BBK1}).
The EUV observations suggest that rich clusters generally have
EUV luminosities of $\sim$$10^{44}$ ergs/s, and have spectra which
decline rapidly in going from the EUV to the X-ray band.

\begin{figure}[t]
\vskip2.8truein
% \special{psfile=ic.ps hscale=40 vscale=40 angle=-90
\includegraphics{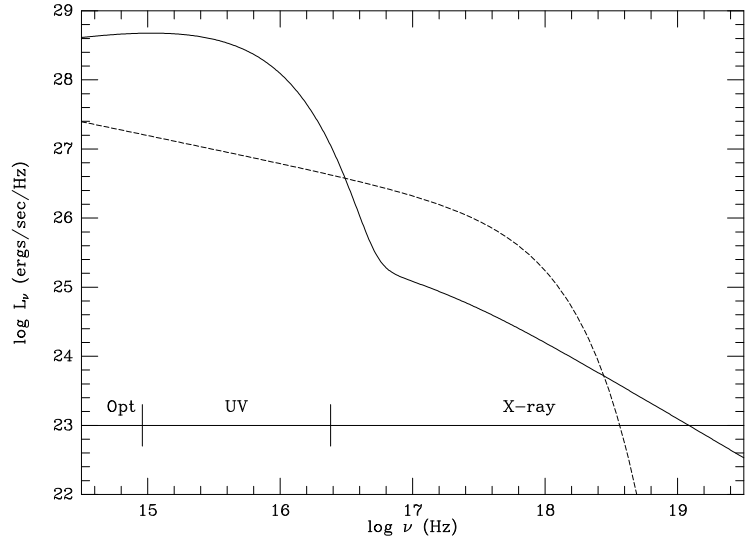}
\caption[]{
The IC spectrum from a typical cluster model (solid curve).
This is the same model as shown in Figure~\protect\ref{fig:espect}.
The dashed curve is a 7 keV thermal bremsstrahlung spectrum.
\label{fig:ic}}
\end{figure}

While it is possible that the EUV emission may be thermal in
origin
(\nolinebreak\cite{F97};
\cite{BLM}),
I believe that it is more likely that this emission is due to
inverse Compton scattering (IC) of CMB photons by low energy
relativistic electrons
(\nolinebreak\cite{H97};
\cite{BB};
\cite{EB};
\cite{SL}).
In this model, the EUV would be produced by electrons with
energies of $\sim$150 MeV ($\gamma \sim 300$; Fig.~\ref{fig:espect}).
As noted above, these electrons have lifetimes which are comparable to
the Hubble time, and should be present in essentially all clusters.
In fact, many of the clusters with observed EUV emission do not appear to
be undergoing mergers at present.
Thus, this emission is not a useful diagnostic for an ongoing merger;
instead, it may represent the emission from electrons accelerated in
many previous mergers.
To produce the EUV luminosities observed, one needs a population of
such electrons with a total energy of $\sim$$10^{62}$ ergs, which is
about 3\% of the typical thermal energy content of clusters.
This is a reasonable acceleration efficiency for these particles, given
that both the thermal energy in the intracluster gas and the relativistic
particles result from merger shocks.
The steep spectrum in going from EUV to X-ray bands is predicted by
this model
(Fig.~\ref{fig:ic});
it results from the rapid increase in losses ($\propto \gamma^2$)
for particles as the energy increases above $\gamma \sim 300$
(Figs.~\ref{fig:losses} \& \ref{fig:lifetime}).

\subsubsection{Hard X-ray Tails} \label{sec:nonthermal_emit_hxr}

If clusters contain higher energy relativistic electrons with
$\gamma \sim 10^4$, these particles will produce hard X-ray emission
by IC scattering.
These are essentially the same electrons which produce the observed
radio halos and relics (\S~\ref{sec:nonthermal_emit_radio}), although
the detailed correspondence depends on the value of the magnetic field.
The ratio of hard X-ray IC emission to radio synchrotron emission allows
one to determine the magnetic field in clusters
(e.g., 
\cite{rephaeli79};
\cite{FFea1}).
Since these higher energy electrons have short lifetimes, they should
only be present in clusters with evidence for a recent or ongoing merger.

Because of the short lifetimes of the electrons producing HXR IC emission,
the population of these particles should be close to steady-state.
If the accelerated electrons have a power-law distribution
(eq.~\ref{eq:power}),
the expected steady-state energy spectral index if IC losses dominated would be
$\alpha_{\rm HXR} = - ( \mu + 1 ) / 2$
(\nolinebreak\cite{ginzberg}).
%(Ginzburg \& Syrovatskii 1964)
For $\mu \approx 2.5 - 4$ (the values expected for typical merger shock
compressions), this gives
$\alpha_{\rm HXR} \approx -1.75$ to $-2.5$.
In the numerical models, the best-fit spectral indices from 20 to 100 keV 
are flatter than this,
$\alpha_{\rm HXR} \approx -1.1$, mainly because other loss processes are
important at the lower energy end of the HXR band
(Fig.~\ref{fig:losses}).

If the population of high energy electrons is in steady state,
the HXR luminosity is just proportional to the energy input from the
mergers into high energy electrons.
To a good approximation, the present day value of $L_{\rm HXR}$
(20--100 keV) is simply given by
\begin{equation} \label{eq:hxr_lum}
L_{\rm HXR} \approx 0.17 {\dot{E}}_{\rm CR,e} ( \gamma > 5000 ) \, .
\end{equation}
where ${\dot{E}}_{\rm CR,e} ( \gamma > 5000 )$ is the total present rate of
injection of energy in cosmic ray electrons with $\gamma > 5000$.
The best-fit coefficient (0.17 in eqn.~\ref{eq:hxr_lum}) depends somewhat
on the power-law index of the injected electrons; the value of 0.17 applies
for $\mu = 2.3$.
Assuming a fixed efficiency
$\epsilon_{\rm CR,e} ( \gamma > 5000 )$
of conversion of shock energy into high energy electrons, 
the rate of particle acceleration is given by
\begin{equation} \label{eq:hxr_lum2}
{\dot{E}}_{\rm CR,e} ( \gamma > 5000 ) =
\epsilon_{\rm CR,e} ( \gamma > 5000 ) \,
{\dot{E}}_{s} \, ,
\end{equation}
where ${\dot{E}}_{s}$ is the total rate of merger shock energy dissipation.
This gives $L_{\rm HXR} \propto {\dot{E}}_{s}$.

Hard X-ray emission in excess of the thermal emission and detected as
a nonthermal tail at energies $\ga$20 keV has been seen in at least
two clusters.
The Coma cluster, which is undergoing at least one merger and which
has a radio halo, was detected with both BeppoSAX and
RXTE
(\nolinebreak\cite{FFea1};
\cite{RGB}).
BeppoSAX has also detected Abell~2256
(\nolinebreak\cite{FFea2}),
another merger cluster with strong diffuse radio emission.
BeppoSAX may have detected Abell~2199
(\nolinebreak\cite{Kea}),
although I believe the evidence is less compelling for this case.
A nonthermal hard X-ray detection of Abell~2199 would be surprising,
as this cluster is very relaxed and has no radio halo or relic
(\nolinebreak\cite{KS1}).

An alternative explanation of the hard X-ray tails is that they might be
due to nonthermal bremsstrahlung
(\nolinebreak\cite{Blasi00};
\cite{dogiel};
\cite{SK}),
which is bremsstrahlung from nonthermal electrons with energies of
10--1000 keV which are being accelerated to higher energies.
The nonthermal tail on the particle distribution might also be associated
with shock acceleration.
On the other hand, these suprathermal electrons have relatively short
time scales to relax into the thermal distribution as a result of
Coulomb collisions.
In fact, this is a general problem of the injection of thermal electrons
into the shock acceleration region.
IC emission from high energy electrons dominates unless the particle
spectrum is very steep
(\nolinebreak\cite{SK}).

The previous hard X-ray detections of clusters have been done with 
instruments with very poor angular resolution.
Thus, they provide no information on the distribution of the hard X-ray
emission.
It would be very useful to determine if the hard X-ray emission is
localized to the radio emitting regions in clusters.
For clusters with radio relics, these might be associated with the
positions of merger shocks in the X-ray images.
Better angular resolution would also insure that the hard X-ray detections
of clusters are not contaminated by emission from other sources.
The IBIS instrument on INTEGRAL will provide a hard X-ray capability with
better angular resolution, and may allow the hard X-ray emission regions
to be imaged
(\nolinebreak\cite{goldoni}).

The predicted IC emission from nonthermal particles is much weaker than
the thermal emission in the central portion of the X-ray band
from about 0.3 keV to 20 keV
(Fig.~\ref{fig:ic}).
However, if the IC emission is localized to merger shock regions,
its local surface brightness might be comparable to the thermal
X-ray emission.
A possible detection of localized IC emission associated with merger shocks
and radio relics has been claimed in Abell~85
(\nolinebreak\cite{bagchi}).
It is possible that Chandra and XMM/Newton will find IC emission
associated with other merger shocks and radio relics.

\subsubsection{Predicted Gamma-Ray and Neutrino Emission}
\label{sec:nonthermal_emit_gamma}

Relativistic electrons and ions in clusters are also expected to produce
strong gamma-ray emission
(\nolinebreak\cite{DS};
\cite{BBP};
\cite{CB};
\cite{BC};
\cite{Blasi99};
\cite{S2})
The region near 100 MeV is particularly
interesting, as this region includes bremsstrahlung from the most common
electrons with $\gamma \sim 300$, and $\pi^o$ decay gamma-rays from ions.
The $\pi^o$ emission mechanism starts with the essentially the same
ion-ion collisions as make secondary electrons
(eq.~\ref{eq:secondary})
\begin{eqnarray}
p + p & \rightarrow & \pi^o + X \, , \nonumber \\
\pi^o & \rightarrow & 2 \gamma \, .
\label{eq:pio}
\end{eqnarray}
Both bremsstrahlung and the $\pi^o$ decay process involve collisions between
relativistic particles (electrons for bremsstrahlung,
ions for $\pi^o$ emission)
and thermal particles, so they should both vary in the same way with
density in the cluster.
Thus, the ratio of these two spectrally distinguishable emission processes
should tell us the ratio of cosmic ray ions to electrons in
clusters
(\nolinebreak\cite{Blasi99};
\cite{S2}).

\begin{figure}[t]
\vskip2.8truein
% \special{psfile=gamma.ps hscale=40 vscale=40 angle=-90
\includegraphics{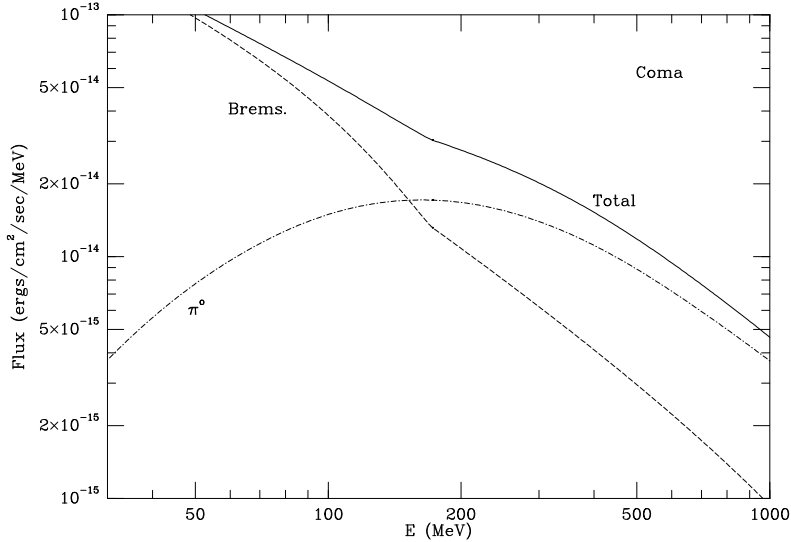}
\caption[]{
The predicted gamma-ray spectrum for the Coma cluster, including
electron bremsstrahlung and $\pi^o$ decay from ions
(\nolinebreak\cite{S2}).
\label{fig:gamma}}
\end{figure}

Figure~\ref{fig:gamma} shows the predicted gamma-ray spectrum for the
Coma cluster, based on a model which reproduces the observed EUV, hard
X-ray, and radio emission
(\nolinebreak\cite{S2}).
The observed upper limit from CGO/EGRET is $<$$4 \times 10^{-8}$
cts/cm$^2$/s for $E > 100$ MeV
(\nolinebreak\cite{Sea}),
while the predicted value for this model is
$\sim$$2 \times 10^{-8}$ cts/cm$^2$/s.
The EGRET upper limit already shows that the ratio of ions to electrons
cannot be too large ($\la$30;
\cite{Blasi99};
\cite{S2}).
The predicted fluxes are such that many nearby clusters should be easily
detectable with GLAST.

The same relativistic particles will also produce neutrinos, which might
be detectable with future instruments
(\nolinebreak\cite{DS};
\cite{BBP};
\cite{CB}).

\subsubsection{Ultra High Energy Cosmic Rays}
\label{sec:nonthermal_emit_cr}

The time scale for most relativistic particles to diffuse out 
of clusters is longer than the Hubble time
(eq.~\ref{eq:diff_time}).
However, very high energy cosmic rays ($E \ga 10^{15}$ eV)
could escape from clusters on relatively short time scales.
In the cosmic ray spectrum seen at the Earth, it is believed that
particles with energies up to $\sim$$10^{14}$ eV come from
supernova explosions in our Galaxy.
Other Galactic sources may produce even higher energy cosmic rays.
However, it is likely that the highest energy cosmic ray particles
($E \ga 3 \times 10^{18}$ eV) are extragalactic in origin
(\nolinebreak\cite{cocconi}).
Merger or accretion shocks in clusters of galaxies are a possible source
of such particles
(e.g., \cite{kang96},1997;
\cite{SO}).
\nocite{kang97}
The advantages of merger shocks are
their high total energies (which helps with the overall flux of cosmic rays), 
their very large physical sizes (which help with the acceleration of
high energy particles with large Larmor radii),
their long time scales
(which helps to provide enough time for the particles to diffuse to these
high energies),
and the relatively low losses in the cluster environment
(\S~\ref{sec:nonthermal_life}).
The Larmor or gyro radius of a high energy particle with a charge $Z$
in the ICM is
\begin{equation} \label{eq:gyro}
r_g = \frac{pc}{Z e B} \approx \frac{0.1}{Z} 
\left( \frac{E}{10^{20} \, {\rm eV}} \right)
\left( \frac{B}{1 \, \mu{\rm G}} \right)^{-1} \, {\rm Mpc} \, ,
\end{equation}
and cluster shock regions are likely to be about this size or larger.
Assuming Bohm diffusion and a strong shock at a velocity $v_s$,
the acceleration time is about
(\nolinebreak\cite{kang96})
\begin{equation} \label{eq:uhe}
t_{\rm acc} \approx 9 \times 10^9 
\left( \frac{E}{10^{20} \, {\rm eV}} \right)
\left( \frac{B}{1 \, \mu{\rm G}} \right)^{-1}
\left( \frac{v_s}{3000 \, {\rm km/s}} \right)^{-2}
\, {\rm yr} \, .
\end{equation}
Thus, it might be possible to accelerate protons up to $\la$$10^{20}$
eV in cluster shocks.

\section{Summary} \label{sec:summary}

I've tried to summarize some of the basic aspects of the physics
of cluster mergers.
Simple estimates for the rates of mergers and for the infall velocities
and impact parameters were given in
\S~\ref{sec:basic}.
The thermal effects of merger shocks are discussed in
\S~\ref{sec:thermal},
with an emphasis on the diagnostics for determining the kinematics of
mergers from X-ray observations of temperatures and densities in
the ICM.
The interaction of cooling flow cores with 
hotter, more diffuse intracluster gas was considered in
(\S~\ref{sec:cf}),
including the mechanism for the disruption of the cooling flow cores
(\S~\ref{sec:cf_cflow}), and the hydrodynamics of
``cold fronts''
(\S~\ref{sec:cf_coldfront}).
Relativistic particles may be accelerated or reaccelerated in merger
shocks or turbulence generated by mergers.
The nonthermal effects of mergers are discussed in
\S~\ref{sec:nonthermal}, including the resulting radio, extreme ultraviolet,
hard X-ray, and gamma-ray emission.

\begin{acknowledgments}
I want to thank my collaborators Josh Kempner, Maxim Markevitch, Scott Randall,
Paul Ricker, and Alexey Vikhlinin for all their help.
I would like to particularly thank Paul Ricker for useful discussions,
and Josh Kempner, Scott Randall, and Yutaka Fujita for careful readings
of a draft.
Scott Randall and Paul Ricker kindly provided figures for this paper.
Support for this work was provided by the National Aeronautics and Space
Administration through Chandra Award Numbers
GO0-1119X, GO0-1173X, GO0-1158X, and GO1-2122X,
issued by the Chandra X-ray Observatory Center, which is operated by the
Smithsonian Astrophysical Observatory for and on behalf of NASA under contract
NAS8-39073.
Support also came from NASA XMM grants
NAG 5-10074 and NAG 5-10075.
\end{acknowledgments}

\end{document}